\documentclass[
    aps,
    pra,
    reprint,
    superscriptaddress
]{revtex4-2}

\usepackage[T1]{fontenc}
\usepackage{amsmath}
\usepackage{amssymb}
\usepackage{amsfonts}
\usepackage{bm}
\usepackage{booktabs}
\usepackage{graphicx}
\usepackage{dcolumn}
\usepackage{multirow}
\usepackage{hyperref}

\newcommand{\bra}[1]{\left\langle #1 \right|}
\newcommand{\ket}[1]{\left| #1 \right\rangle}

\begin{document}

\title{CSS codes for Quantum Metrology with Discrete-time Error Correction}

\author{Ugnė Liaubaitė}
\affiliation{Institute of Data Science and Digital Technologies, Faculty of Mathematics and Informatics, Vilnius University}

\author{Debora Ramacciotti}
\author{Robert Raußendorf}
\affiliation{Institute for Theoretical Physics, Leibniz University Hannover}

\date{\today}

\begin{abstract}

The goal of quantum metrology is to estimate an unknown parameter with
better-than-classical precision and, ideally, to attain Heisenberg
scaling. In realistic settings, however, noise can substantially reduce
this advantage and, in many cases, restore standard-quantum-limit-like
performance. Preserving the quantum enhancement therefore requires
noise-robust strategies, which can be implemented using quantum error
correction. In this work, we develop and analyze a quantum-metrology protocol based on Calderbank--Shor--Steane (CSS) codes. The main result is that, for noise perpendicular to the sensing Hamiltonian, discrete-time error correction
preserves Heisenberg-like temporal scaling over a finite
interrogation-time window whose duration depends on the correction
frequency. It is further shown that recovery applied only
after the sensing evolution provides no metrological advantage over the
corresponding uncorrected noisy protocol in the setup considered here. An analytical framework is developed for arbitrary CSS codes and applied to the Steane and Shor
codes, for which the discrete-time corrected dynamics, the
Heisenberg-like-to-SQL-like crossover, and the optimal interrogation time
are determined and numerically analysed. 

\end{abstract}

\maketitle

\section{Introduction}

Quantum metrology is concerned with the estimation of unknown parameters
encoded in quantum states with precision exceeding that achievable using
classical resources. A central question is how the estimation uncertainty
scales with the available physical resources. For $N$ independent
parallel probes, the standard quantum limit (SQL) scales as $N^{-1/2}$, whereas
Heisenberg scaling corresponds to $N^{-1}$ \cite{Giovannetti2011}. In the temporal setting, the analogous distinction appears in the scaling with the total sensing time. Quantum-enhanced strategies attaining the latter scaling are generally more vulnerable to noise, which can change the asymptotic scaling from inverse-linear to inverse-square-root. We therefore refer to \(\delta\omega\propto T^{-1}\) as Heisenberg-like temporal scaling and to \(\delta\omega\propto T^{-1/2}\) as SQL-like temporal scaling.

Quantum-enhanced sensing has been
investigated extensively both theoretically
\cite{Giovannetti2011,Boixo_2007,Pezz2018} and experimentally
\cite{Slusher1985,Leibfried2004,Hosten2016}, including magnetic-field
sensing using NOON/GHZ-type states \cite{Jones2009}.

Noise constitutes a major limitation to quantum-enhanced metrology. In
the presence of generic noise, the advantage provided by entanglement can
be strongly reduced and, asymptotically, the achievable scaling may revert
to the SQL \cite{Huelga1997,Demkowicz2012}. Different approaches have
therefore been investigated for preserving metrological sensitivity in
noisy systems, including squeezed states \cite{Hosten2016},
decoherence-free subspaces \cite{Jeske2014}, and quantum error correction
(QEC) \cite{Zhou2018,Dur2014,Arrad2014,Kessler2014}.

Quantum error correction was originally developed in the context of
quantum information processing and has subsequently been incorporated
into quantum-metrological protocols. In this setting, the usual
error-correction conditions must be supplemented by the requirement that
the parameter-dependent signal is preserved. A necessary and sufficient
condition for the attainability of Heisenberg scaling under Markovian
noise was derived in Ref.~\cite{Zhou2018} and is known as the
Hamiltonian-not-in-Lindblad-span (HNLS) condition. Under the assumptions
considered therein, Heisenberg scaling can be achieved by an
error-corrected protocol if and only if the signal Hamiltonian is not
contained in the Lindblad span, describing the noise.

Error-corrected quantum metrology has been investigated for different
sensing Hamiltonians, noise models, and QEC constructions
\cite{Dur2014,Arrad2014,Kessler2014, Rojkov2022}. Most theoretical treatments of the
HNLS-compatible setting consider the limit of arbitrarily frequent or
sufficiently fast correction, while finite-frequency or finite-strength
QEC has also been studied within specific sensing and correction models
\cite{Shettell_2021, Hu2022}. A general analytical characterization of stabilizer-code protocols under discrete-time correction, including the resulting transition between Heisenberg-like and SQL-like temporal scaling, therefore remains of interest.

In this work, a quantum-metrological protocol based on
Calderbank--Shor--Steane (CSS) codes is constructed. Separation of
 stabilizer generators into strictly $X$- and $Z$-type operators allows a
partial error-correction scheme to be implemented. For noise perpendicular to the signal Hamiltonian, one CSS stabilizer sector is used for syndrome extraction and recovery, while the complementary sector is used for the metrological readout. Noise parallel to the signal is
considered separately as the HNLS-incompatible case. 

The discrete-time regime is investigated by retaining a finite correction
interval $\tau$ between consecutive recovery operations. Within this protocol, it is shown that
discrete-time error correction is sufficient to preserve Heisenberg-like
temporal scaling over a nonzero interrogation-time window whose
duration increases as the correction interval is reduced. However, for any fixed
nonzero $\tau$, the remaining decay eventually leads to a finite
optimal interrogation time and a crossover to SQL-like scaling, thereby
excluding asymptotic Heisenberg-like scaling. 

In addition, a recovery only at the end of the protocol is shown to provide no metrological advantage within the
proposed protocol.

Analytical expressions are obtained for arbitrary CSS codes, and the
framework is applied explicitly to the seven-qubit Steane code and the
nine-qubit Shor code. For both examples, the performance of the encoded probe is compared with that of the GHZ and product-state probes. Further, the discrete-time corrected dynamics are characterized and numerically analyzed, and the Heisenberg-like-to-SQL-like crossover and optimal interrogation time are determined.

\section{Theory}
\label{sec:theory}

The sensing Hamiltonian considered throughout this work is a local
$Z$ Hamiltonian,
\begin{equation}
    H
    =
    \omega \sum_{i=1}^{n} Z_i,
    \label{eq:hamiltonian}
\end{equation}
acting on $n$ physical qubits, where $\omega$ denotes the unknown
parameter to be estimated. Such a Hamiltonian can, for example, describe
the coupling of $n$ spins to a uniform magnetic field, with $\omega$
proportional to the corresponding field strength.

Noise parallel to the signal Hamiltonian is described by
$L_i=\sqrt{\gamma}\,Z_i$, whereas perpendicular noise is described by
$L_i=\sqrt{\gamma}\,X_i$. For the Hamiltonian in
Eq.~\eqref{eq:hamiltonian}, the HNLS condition is satisfied for
perpendicular $X$ noise but not for parallel $Z$ noise. Consequently,
error correction aimed at preserving Heisenberg-like scaling is applied
in the perpendicular-noise case, while the parallel-noise case is
considered without error correction.

The noisy dynamics under Markovian noise is described by a Lindblad master equation. For
convenience of analytical treatment, the Heisenberg picture is used. An observable $O$ evolves according to the adjoint Lindblad equation
\begin{equation}
    \frac{dO}{dt}
    =
    i[H,O]
    +
    \sum_j
    \left(
        L_j^\dagger O L_j
        -
        \frac{1}{2}
        \left\{
            L_j^\dagger L_j,O
        \right\}
    \right).
    \label{eq:Lindblad_Heisenberg}
\end{equation}

A general $[[n,k,d]]$ CSS code
\cite{Calderbank1996,Steane1996} is considered, encoding $k$ logical
qubits into $n$ physical qubits. The encoded probe is initialized in the
logical-plus state of an $[[n,k,d]]$ CSS code
\begin{equation}
    \ket{\psi_0}
    =
    \ket{+_L}^{\otimes k}.
    \label{eq:initial_state}
\end{equation}
This state is chosen for convenience, since in the idealized noiseless scenario
the quantum Fisher information is independent of the choice of pure logical state
for stabilizer codes with distance $d\geq 3$. In the presence of noise, however,
the optimal encoded initial state is generally nontrivial and is not optimized here.

The measurement considered in this work is the projector onto the joint
$+1$ eigenspace of all $X$-type stabilizers of the CSS code,
\begin{equation}
    \Pi_X
    =
    \prod_{j=1}^{r_X}
    \frac{I+S_j^X}{2}
    =
    \frac{1}{|G_X|}
    \sum_{g\in G_X} g,
    \label{eq:projector}
\end{equation}
where $\{S_j^X\}_{j=1}^{r_X}$ is a set of independent generators of the
$X$-stabilizer group $G_X$, with $r_X=\log_2|G_X|$. Its expectation value,
\begin{equation}
    P_X(\omega,t)
    =
    \operatorname{Tr}\!\left[\rho_0\Pi_X(t)\right]
    =
    \frac{1}{|G_X|}
    \left(
        1+
        \sum_{g\in G_X\setminus\{I\}}
        \langle g(t)\rangle
    \right),
    \label{eq:exp_val_Px}
\end{equation}
gives the probability that all $X$-type stabilizer measurements yield
the outcome $+1$. The corresponding reduction to individual stabilizer
expectation values is derived in
Appendix~\ref{app:projector_probability}.

Since $\Pi_X$ is a binary projective measurement, its variance is
\begin{equation}
    (\Delta\Pi_X)^2
    =
    P_X(\omega,t)
    \left[
        1-P_X(\omega,t)
    \right].
    \label{eq:projector_variance}
\end{equation}

For the corresponding binary measurement, the classical Fisher
information\footnote{Since the measurement is fixed to $\Pi_X$, the
relevant quantity is the classical Fisher information associated with
the resulting binary outcome distribution; no optimization over
measurements is performed.}
is
\begin{equation}
    F_{\mathrm{C}}(\omega,t)
    =
    \frac{
        \left[
            \partial_\omega P_X(\omega,t)
        \right]^2
    }{
        P_X(\omega,t)
        \left[
            1-P_X(\omega,t)
        \right]
    }.
    \label{eq:classical_FI}
\end{equation}

After $R_{\mathrm{rep}}$ independent repetitions of the protocol, the
corresponding uncertainty obtained through error propagation is
\begin{equation}
    \delta\omega(t)
    =
    \frac{
        \sqrt{
            P_X(\omega,t)
            \left[
                1-P_X(\omega,t)
            \right]
        }
    }{
        \sqrt{R_{\mathrm{rep}}}\,
        \left|
            \partial_\omega P_X(\omega,t)
        \right|
    }.
    \label{eq:deltaomega}
\end{equation}

In the following, the metrological performance of the protocol is
characterized through the inverse uncertainty,
\begin{equation}
    \delta\omega^{-1}(t)
    =
    \sqrt{
        R_{\mathrm{rep}}
        F_{\mathrm{C}}(\omega,t)
    },
    \label{eq:sensitivity}
\end{equation}
which will be referred to as the sensitivity. Its dependence on the
sensing time is used below to characterize the Heisenberg-like and
standard-quantum-limit-like scaling regimes.

The single-qubit Lindblad dynamics can be solved analytically for
both parallel and perpendicular noise. The resulting evolution is
expressed in terms of the functions $a_\nu(t)$, $b_\nu(t)$, and
$c_\nu(t)$, with $\nu\in\{X,Z\}$ denoting the noise model. Their explicit
forms and frequency derivatives are derived in
Appendix~\ref{app:single_qubit_heisenberg}. These solutions are then
extended to an arbitrary $X$-type stabilizer in
Appendix~\ref{app:noisy_X_stabilizer} and, consequently, to the full
projector $\Pi_X$ in Appendix~\ref{app:projector_probability}. The
resulting stabilizer and projector expectation values, together with
their frequency derivatives, provide the analytical quantities used
throughout the remainder of the work.

\section{Methods}
\label{sec:methods}

\subsection{Parameter estimation}

Each realization of the sensing protocol consists of preparing the
initial state in Eq.~\eqref{eq:initial_state}, evolving the system for an
interrogation time $t$, applying discrete error-correction steps at fixed
intervals $\tau$, and finally measuring all $X$-type stabilizers. The
final measurement yields a binary outcome from which
$P_X(\omega,t)=\langle\Pi_X\rangle$ is reconstructed. Repeating the
protocol $R_{\mathrm{rep}}$ times provides the statistics used for
parameter estimation.
Note that in case of parallel noise, error correction is not applied, and that step is skipped.

When the analytical dependence of $P_X(\omega,t)$ on $\omega$ can be
inverted, an inversion estimator can be constructed from the empirical
probability
\begin{equation}
    \widehat{P}_X
    =
    \frac{R_{\mathrm{rep}}^{+}}{R_{\mathrm{rep}}},
\end{equation}
such that, on the appropriate branch,
\begin{equation}
    \widehat{\omega}
    =
    P_X^{-1}(\widehat{P}_X,t).
\end{equation}
The relevant branch may be fixed using prior information or a preliminary
estimate of $\omega$.

Even in a case where a closed-form inversion is not available, a maximum-likelihood
estimator may instead be employed, which allows a very general application of the protocol. Given
$R_{\mathrm{rep}}^{+}$ outcomes in the joint $+1$ eigenspace of the
$X$-type stabilizers, the likelihood is
\begin{equation}
    \mathcal{L}(\omega)
    =
    \binom{R_{\mathrm{rep}}}{R_{\mathrm{rep}}^{+}}
    \left[P_X(\omega,t)\right]^{R_{\mathrm{rep}}^{+}}
    \left[1-P_X(\omega,t)\right]^{
        R_{\mathrm{rep}}-R_{\mathrm{rep}}^{+}
    }.
    \label{eq:likelihood}
\end{equation}

\subsection{Discrete-time error correction}

Error correction is applied for perpendicular $X$ noise using the
$Z$-type stabilizer syndromes. Recovery operations are applied at fixed
time intervals $\tau$ and are assumed to be instantaneous and ideal. The
system evolves according to Eq.~\eqref{eq:Lindblad_Heisenberg} between
successive recovery operations.

In the Heisenberg picture, both the noisy evolution and the recovery of
an $X$-type stabilizer $g$ preserve the operator space spanned by
operators of the form $gZ(u)$. The dynamics can therefore be reduced to
the evolution of their coefficients. The complete construction of the
recovery matrix, the noisy evolution matrix, and the corresponding
frequency derivatives is given in
Appendix~\ref{app:corrected_evolution}.

The time-evolved projector probability is given by Eq.~\eqref{eq:exp_val_Px}. The expectation value of an $X$-type stabilizer at a correction time $t=M\tau$, after $M$ complete correction cycles, is
\begin{equation}
    \langle g(t)\rangle_M
    =
    \sum_{\substack{
        u\in\mathbb{F}_2^n\\
        Z(u)\in G_Z
    }}
    \left[
        \left(
            E^{(g)}(\omega,\tau)R
        \right)^M
        e_0
    \right]_u ,
    \label{eq:corrected_stabilizer_general}
\end{equation}
where $e_0$ denotes the coefficient vector corresponding to $u=0$, $E^{(g)}$ is the evolution operator for the coefficients and $R$ is the recovery operator, defined explicitly in Appendix~\ref{app:corrected_evolution}.

The extension to an arbitrary running time $t$, including a final
incomplete correction interval, is also given in
Appendix~\ref{app:corrected_evolution}.

\subsection{Sensitivity analysis and optimal interrogation time}

The metrological performance is characterized by the inverse uncertainty
$\delta\omega^{-1}$ defined in Eq.~\eqref{eq:sensitivity}. The projector probability and, consequently, the sensitivity oscillate in time. The local maxima of the sensitivity are therefore used to define an envelope that characterizes the metrological performance.

At the correction times $t=M\tau$, the projector probability admits the
spectral representation which can be effectively derived analytically using the Krylov-space reduction. The full reduction and spectral construction are given in Appendix~\ref{app:spectral_reduction}. The final result for the spectral representation of the projector probability is given by: 
\begin{equation}
    P_X(\omega,t)
    =
    P_{1}
    +
    \sum_r A_r e^{-D_r t}
    +
    \sum_k B_k e^{-D_k t}
    \cos(\Omega_k t+\phi_k),
    \label{eq:PX_spectral_form}
\end{equation}
where $P_1$ is the contribution from stationary modes with
$\lambda=1$, $A_r$ are the amplitudes of real non-oscillating decaying
modes, and $B_k$ are the amplitudes of oscillating complex-conjugate mode
pairs. The rates $D_r$ describe decay of the real modes, whereas $D_k$
describe decay of the oscillating modes. Both are obtained from the
corresponding eigenvalue modulus,
$D_j=-\tau^{-1}\log|\lambda_j|$, while
$\Omega_k=\arg(\lambda_k)/\tau$ gives the oscillation frequency.

For sufficiently small correction intervals, the sensitivity envelope is
well described by a two-factor decay containing contributions proportional
to $\sqrt{\tau t}$ and $\tau t$. The corresponding coefficients are
code dependent and are determined explicitly for the Steane and Shor
codes in Appendix~\ref{app:code_specific_results}.

To determine the optimal allocation of a fixed total sensing-time budget $T$ between interrogation duration and repeated measurements, we consider interrogations of duration $t$. An interrogation of
duration $t$ can be repeated $\lfloor T/t\rfloor$ times within this
budget, giving the accumulated sensitivity
\begin{equation}
    \Sigma(t;T)
    =
    \sqrt{
        \left\lfloor
            \frac{T}{t}
        \right\rfloor
    }\,
    \delta\omega_{\mathrm{env}}^{-1}(t).
    \label{eq:repeated_sensitivity}
\end{equation}
The optimal interrogation time is therefore defined as
\begin{equation}
    t_{\mathrm{opt}}
    =
    \underset{0<t\leq T}{\arg\max}\;
    \Sigma(t;T).
    \label{eq:topt}
\end{equation}
Beyond $t_{\mathrm{opt}}$, the interrogation time is kept fixed, and the
protocol is repeated. The resulting crossover in the accumulated
sensitivity is analyzed in Sec.~\ref{sec:results}.

\section{Results}
\label{sec:results}

The general theoretical framework and discrete-time error-correction
procedure were introduced in the preceding sections. Here, the analytical
and numerical results are presented for two representative CSS codes:
the seven-qubit Steane code and the nine-qubit Shor code. The detailed
code-specific analytical derivations are given in
Appendices~\ref{app:steane_analytic} and \ref{app:shor_analytic},
respectively.

Since the number of repetitions $R_{\mathrm{rep}}$ contributes only an
overall factor $\sqrt{R_{\mathrm{rep}}}$ to the inverse uncertainty, the
comparisons below are presented for $R_{\mathrm{rep}}=1$.

In figures displaying sensitivity envelopes, the markers denote local maxima obtained numerically from the analytically derived sensitivity, while the corresponding lines show envelope fits to these maxima.

\subsection{Parallel noise without error correction}
\label{sec:results_Z_noise}

We first consider noise parallel to the sensing Hamiltonian,
$L_i=\sqrt{\gamma}\,Z_i$. Since the HNLS condition is not satisfied in
this case, no error correction is applied.

To obtain the sensitivity envelopes, define
$r_{\mathrm{St}}= e^{-8\gamma t}$ and 
$r_{\mathrm{Sh}} = e^{-12\gamma t}$
which yields the following analytical expressions for the encoded Steane
and Shor probes:
\begin{widetext}
\begin{equation}
    \delta\omega_{\mathrm{St,env}}^{-1}(t)
    =
    \frac{
        4\sqrt{7}\,t\,r_{\mathrm{St}}
    }{
        \sqrt{
            2
            +9r_{\mathrm{St}}
            -7r_{\mathrm{St}}^2
            +
            \sqrt{
                (2-r_{\mathrm{St}})
                (1-r_{\mathrm{St}})
                (7r_{\mathrm{St}}+1)
                (7r_{\mathrm{St}}+2)
            }
        }
    },
    \label{eq:results_Z_steane_envelope}
\end{equation}

\begin{equation}
    \delta\omega_{\mathrm{Sh,env}}^{-1}(t)
    =
    \frac{
        6\sqrt{6}\,t\,r_{\mathrm{Sh}}
    }{
        \sqrt{
            1+r_{\mathrm{Sh}}
            +
            \sqrt{
                (1-r_{\mathrm{Sh}})
                (3r_{\mathrm{Sh}}+1)
            }
        }
    }.
    \label{eq:results_Z_shor_envelope}
\end{equation}
\end{widetext}

\begin{figure*}[t]
    \centering
    \includegraphics[width=0.95\textwidth]{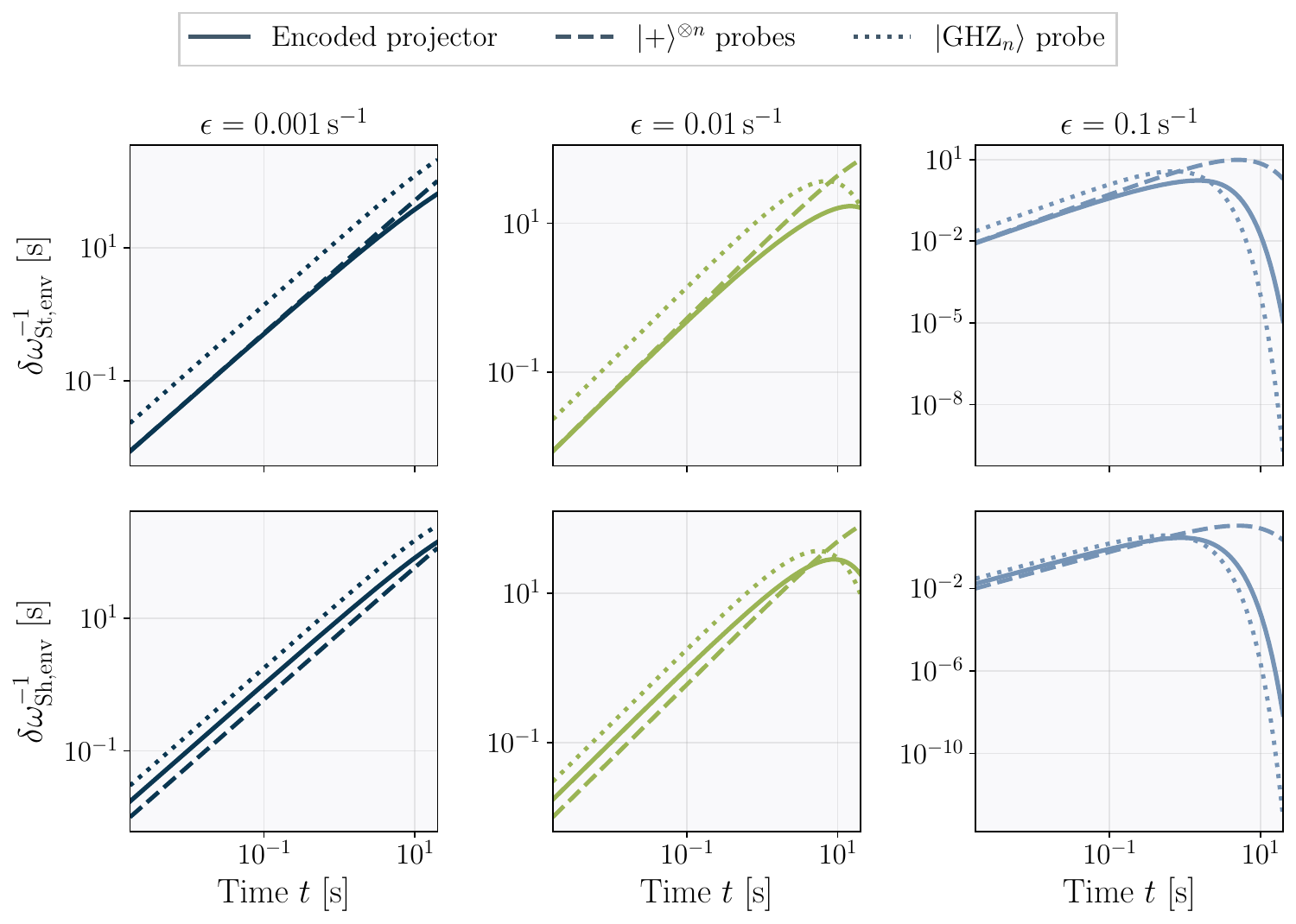}

    \caption{
        Envelope of the frequency sensitivity under uncorrected $Z$
        noise. Rows correspond to the Steane and Shor codes and columns
        to different values of the noise strength $\gamma$. Solid,
        dashed, and dotted lines denote the encoded projector protocol,
        $n$ independent $\ket{+}$ probes, and the $n$-qubit GHZ probe,
        respectively. The notation $\epsilon$ used in the figure
        corresponds to the noise strength, $\epsilon=\gamma$.
    }
    \label{fig:Z_noise_sensitivity}
\end{figure*}

Figure~\ref{fig:Z_noise_sensitivity} compares the uncorrected encoded
probes with the $n$-qubit GHZ state, $\ket{\mathrm{GHZ}_n}=(\ket{0}^{\otimes n}+\ket{1}^{\otimes n})/\sqrt{2}$, and the corresponding product-state
probe, $\ket{\psi_{\mathrm{prod}}}=\ket{+}^{\otimes n}$. 
All reference implementations use the same $X$-basis readout: individual $X$ outcomes are used for the independent probes, while their parity $X^{\otimes n}$ is used for the GHZ probe.
In the weak-noise regime, the GHZ probe provides the highest
sensitivity but also exhibits the fastest decay as the noise strength is
increased. The encoded probes retain their sensitivity slightly longer
than the GHZ probe, but provide no advantage over the independent probes with increasing noise 
when the parallel noise remains uncorrected. At larger noise strengths,
the product-state strategy is therefore the most robust among the
considered probes. This behaviour reflects the trade-off between the
metrological enhancement provided by entanglement and its greater
sensitivity to local noise.

\subsection{Discrete-time correction under perpendicular noise}
\label{sec:results_X_noise}

When considering perpendicular noise,
$L_i=\sqrt{\gamma}\,X_i$, the discrete-time recovery introduced
in Sec.~\ref{sec:methods}. is applied. For the numerical analysis below,
the signal frequency is fixed to $\omega=3$ and the physical noise
strength to $\gamma=0.1$, while the correction interval $\tau$ is varied.
The corresponding code-specific spectral decompositions are derived in
Appendices~\ref{app:steane_analytic} and \ref{app:shor_analytic}.

For the Steane code, the projector probability is
\begin{equation}
    P_{X,\mathrm{St}}(t)
    \simeq
    \frac{1}{8}
    +
    \frac{21}{32}e^{-D_r t}
    +
    \frac{7}{32}e^{-D_o t}
    \cos\!\left(
        \Omega_{\mathrm{St}}t+\phi_{\mathrm{St}}
    \right),
    \label{eq:results_steane_projector}
\end{equation}

For a small correction interval $\tau$, the corresponding decay rates are
\begin{align}
    D_r
    &=
    6\gamma^2\tau
    +
    \left(
        \frac{32}{3}\gamma\omega^2
        -
        32\gamma^3
    \right)\tau^2
    +
    O(\tau^3),
    \label{eq:results_Dr}
    \\
    D_o
    &=
    18\gamma^2\tau
    +
    \left(
        \frac{32}{3}\gamma\omega^2
        -
        92\gamma^3
    \right)\tau^2
    +
    O(\tau^3),
    \label{eq:results_Do}
\end{align}
while only the leading-order oscillation frequency is required here
\begin{equation}
    \Omega_{\mathrm{St}}
    =
    8\omega+O(\tau).
\end{equation}

For the Shor code, the projector probability takes the form
\begin{equation}
    P_{X,\mathrm{Sh}}(t)
    \simeq
    \frac{1}{4}
    +
    \frac{3}{8}e^{-D_{\mathrm{Sh}}t}
    +
    \frac{3}{8}e^{-D_{\mathrm{Sh}}t}
    \cos\!\left(
        \Omega_{\mathrm{Sh}}t+\phi_{\mathrm{Sh}}
    \right),
    \label{eq:results_shor_projector}
\end{equation}
where
\begin{equation}
    D_{\mathrm{Sh}}
    =
    6\gamma^2\tau
    +
    \left(
        16\gamma\omega^2
        -
        16\gamma^3
    \right)\tau^2
    +
    O(\tau^3),
    \label{eq:results_Dsh}
\end{equation}
and
\begin{equation}
    \Omega_{\mathrm{Sh}}
    =
    12\omega+O(\tau).
\end{equation}

For both codes, the leading decay is proportional to
$\gamma^2\tau$ and is therefore suppressed as the correction interval is
reduced.

\begin{figure*}[t]
    \centering

    \includegraphics[width=0.90\textwidth]{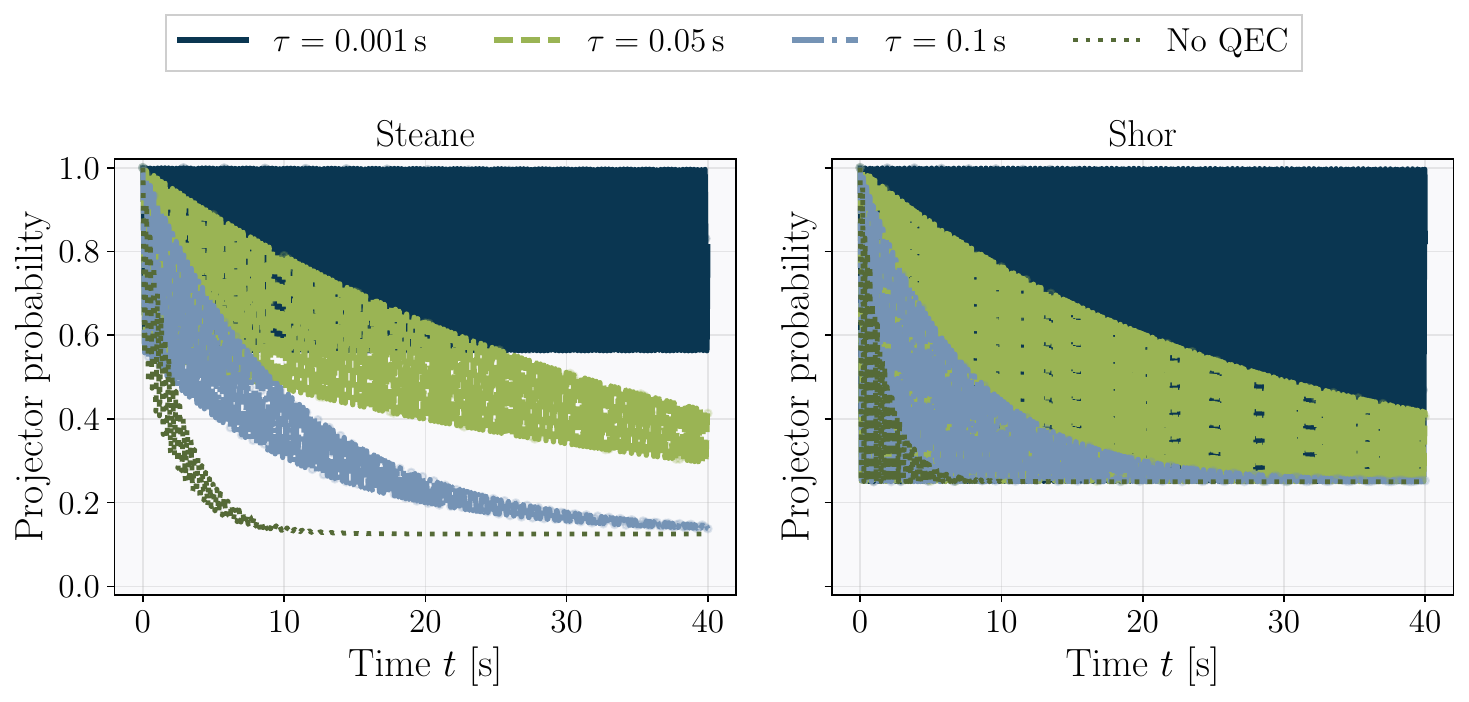}

    \caption{
Projector probability \(P_X(\omega,t)\) under discrete \(X\)-error correction for the Steane code (left) and the Shor code (right), for different correction intervals \(\tau\) at \(\omega=3\). The dense appearance reflects the rapid parameter-dependent oscillations.
    }
    \label{fig:projector_X_QEC}
\end{figure*}

The corresponding projector probabilities are shown in
Fig.~\ref{fig:projector_X_QEC}. As the correction interval is reduced,
the decay of the oscillating signal is increasingly suppressed, and the
dynamics approach the corresponding noiseless limit.

\subsection{Sensitivity under discrete-time correction}
\label{sec:results_sensitivity}

The spectral decay of the projector probability determines the
corresponding sensitivity envelope. For the Steane code,
\begin{equation}
    \delta\omega_{\mathrm{St,env}}^{-1}(t)
    \simeq
    2\sqrt{7}\,t
    \exp\!\left[
        -D_{1/2}^{\mathrm{St}}\sqrt{\tau t}
        -
        D_1^{\mathrm{St}}\tau t
    \right],
    \label{eq:results_steane_envelope}
\end{equation}
where
\begin{align}
    D_{1/2}^{\mathrm{St}}
    &=
    \frac{3}{4\sqrt{2}}
    \sqrt{
        \frac{3D_r+D_o}{\tau}
    },
    \label{eq:results_steane_Dhalf}
    \\
    D_1^{\mathrm{St}}
    &=
    \frac{D_o}{2\tau}.
    \label{eq:results_steane_Done}
\end{align}

For the Shor code,
\begin{equation}
    \delta\omega_{\mathrm{Sh,env}}^{-1}(t)
    \simeq
    6\sqrt{3}\,t
    \exp\!\left[
        -D_{1/2}^{\mathrm{Sh}}\sqrt{\tau t}
        -
        D_1^{\mathrm{Sh}}\tau t
    \right],
    \label{eq:results_shor_envelope}
\end{equation}
with
\begin{align}
    D_{1/2}^{\mathrm{Sh}}
    &=
    \frac{1}{2}
    \sqrt{
        \frac{D_{\mathrm{Sh}}}{\tau}
    },
    \label{eq:results_shor_Dhalf}
    \\
    D_1^{\mathrm{Sh}}
    &=
    \frac{D_{\mathrm{Sh}}}{2\tau}.
    \label{eq:results_shor_Done}
\end{align}

The two-factor envelopes are fitted numerically to the exact local
maxima of the sensitivity.

\begin{figure*}[t]
    \centering

     \includegraphics[width=0.92\textwidth]{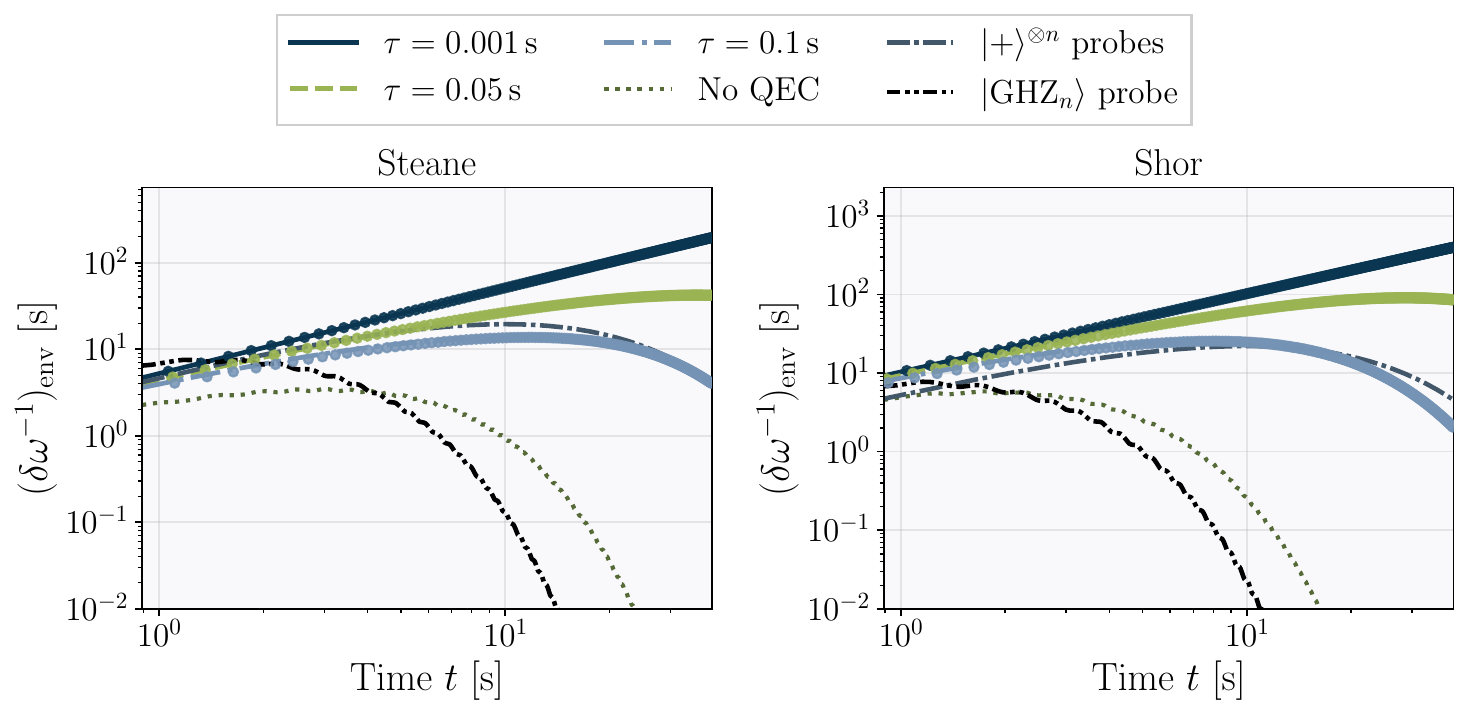}

    \caption{
        Frequency-sensitivity envelopes under discrete $X$-error
        correction for the Steane code (left) and the Shor code (right).
        Colored markers denote the exact local maxima of
        $\delta\omega^{-1}$, while the corresponding continuous curves
        show the fitted two-factor envelope given by Eq.~\eqref{eq:sensitivity_envelope}. The uncorrected encoded
        protocol and the GHZ probe are included as analytical benchmarks
        and are not part of the fit. 
        }
    \label{fig:sensitivity_X_QEC}
\end{figure*}

Figure~\ref{fig:sensitivity_X_QEC} shows that the Shor and Steane codes provide comparable results. The discrete-time corrected encoded
protocols outperform the GHZ benchmark over the considered interval and,
for $\tau=0.1$, become comparable to the independent-probe strategy.
The noiseless encoded benchmark also remains above the GHZ benchmark
over a longer interrogation interval.

\subsection{Optimal interrogation time and HL--SQL crossover}
\label{sec:results_crossover}

The remaining decay at a finite correction interval eventually makes
continued uninterrupted interrogation suboptimal. For the fixed simulated
sensing-time budget $T$, the optimal interrogation time is
determined according to Eq.~\eqref{eq:topt}. In the numerical analysis, the sensing-time budget is considered to be \(T_{\max}=40\,\mathrm{s}\).

For comparison with the numerical optimum, neglecting the integer
rounding in the repetition number gives the continuous approximation
\begin{equation}
    \widetilde{t}_{\mathrm{opt}}
    =
    \min\!\left\{
        \frac{
            4
        }{
            \tau
            \left[
                D_{1/2}
                +
                \sqrt{D_{1/2}^{2}+8D_1}
            \right]^2
        },
        T
    \right\}.
    \label{eq:results_topt_analytic}
\end{equation}

The numerical and analytical values are summarized in Table~\ref{tab:topt}, where \(T=T_{\max}\).

\begin{table*}[t]
    \centering
    \caption{
        Optimal interrogation times for the fixed total simulated budget
        $T$. The numerical optimum maximizes the numerical
        envelope with the exact integer repetition factor
        $\lfloor T/t\rfloor$. The analytical approximation is
        obtained by replacing $\lfloor T/t\rfloor$ by $T/t$ in the
        two-factor model. The floor loss is
        $100\left[
        1-\lfloor T/t_{\mathrm{opt}}\rfloor/
        (T/t_{\mathrm{opt}})
        \right]$.
        If the turnover is not resolved before the end of the simulated
        interval, the table reports $T$.
    }
    \label{tab:topt}
    \vspace{5mm}
    \begin{tabular}{lccccc}
        \toprule
        Code
        & $\tau$ [s]
        & $t_{\mathrm{opt}}$ [s]
        & $\widetilde{t}_{\mathrm{opt}}$ [s]
        & Relative deviation
        & Floor loss [\%]
        \\
        \midrule
        Steane & $10^{-3}$ & 40    & 40    & 0       & 0      \\
        Steane & 0.05      & 13.21 & 12.31 & 0.06873 & 0.8917 \\
        Steane & 0.10      & 5.61  & 3.749 & 0.3318  & 1.817  \\
        Shor   & $10^{-3}$ & 40    & 40    & 0       & 0      \\
        Shor   & 0.05      & 13.20 & 16.26 & 0.2323  & 1.017  \\
        Shor   & 0.10      & 3.567 & 4.541 & 0.2730  & 1.908  \\
        \bottomrule
    \end{tabular}
\end{table*}
    \vspace{5mm}

To display the crossover, we define the accumulated sensitivity
\begin{equation}
    \Sigma_\tau(T;t_{\mathrm{opt}})
    =
    \begin{cases}
        \delta\omega_{\mathrm{env}}^{-1}(T),
        & T\leq t_{\mathrm{opt}},\\[6pt]
        \displaystyle
        \sqrt{\left\lfloor
        \frac{T}{t_{\mathrm{opt}}}
        \right\rfloor}\,
        \delta\omega_{\mathrm{env}}^{-1}(t_{\mathrm{opt}}),
        & T>t_{\mathrm{opt}}.
    \end{cases}
    \label{eq:crossover_sensitivity}
\end{equation}

For any fixed nonzero correction interval, the remaining decay ultimately
produces a finite $t_{\mathrm{opt}}$. Before this turnover, however, the
corrected protocol retains approximately linear-in-time,
Heisenberg-like sensitivity over a finite interval. Beyond
$t_{\mathrm{opt}}$, independent repetitions at fixed interrogation time
give the square-root accumulation characteristic of the
standard-quantum-limit-like regime. The duration of the Heisenberg-like
window increases as the correction interval $\tau$ is reduced.

\begin{figure*}[t]
    \centering

    \includegraphics[width=0.92\textwidth]{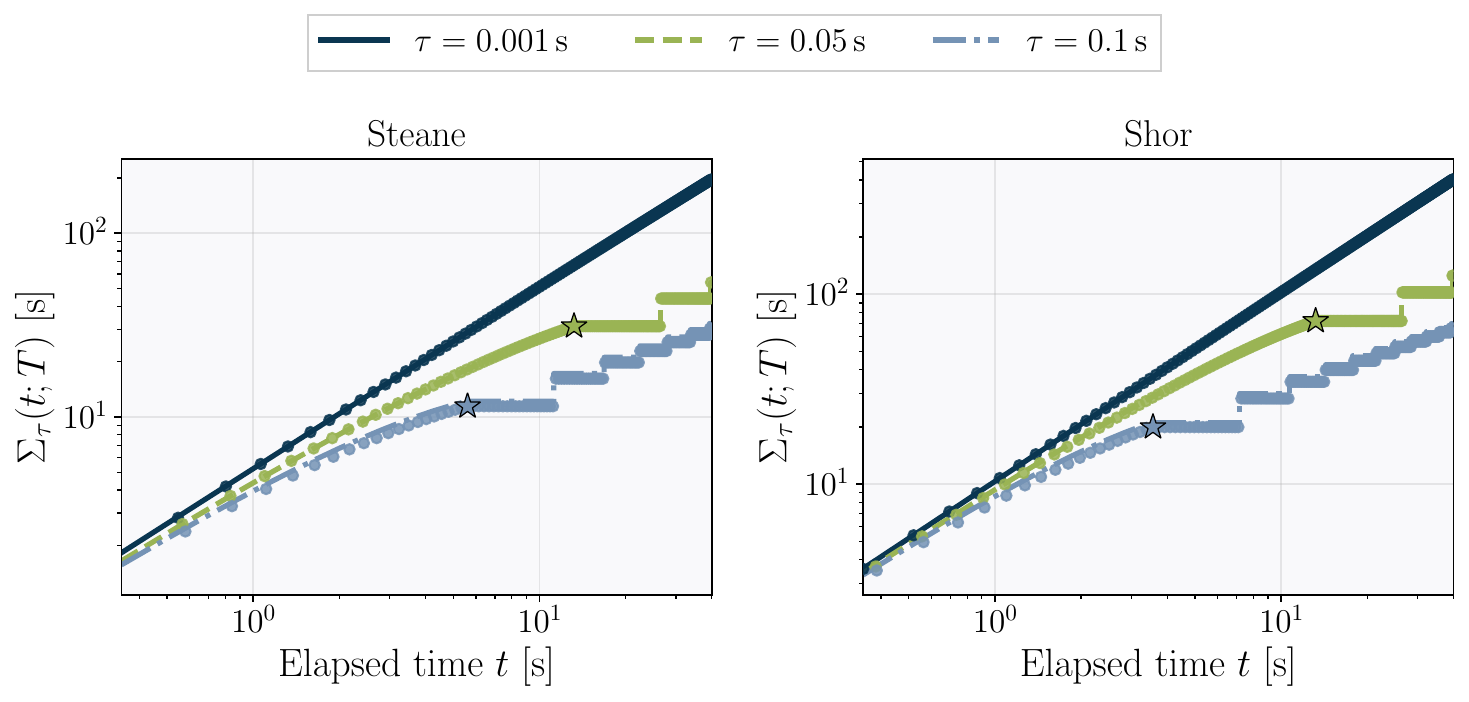}

    \caption{
        Crossover from the Heisenberg-like to the SQL-like regime under
        discrete $X$-error correction for the Steane code (left) and the
        Shor code (right), for different correction intervals $\tau$.
        The interrogation time $t_{\mathrm{opt}}(T)$ is optimized for
        the fixed final budget $T$. Colored dots show the numerical results, while
solid curves show the fitted two-factor envelope.
        Beyond $t_{\mathrm{opt}}$, the protocol is repeated at fixed
        interrogation time, producing the staircase behavior. Stars mark the numerically
determined $t_{opt}(T)$.
    }
    \label{fig:HL_SQL_crossover}
\end{figure*}

\subsection{Rescaled crossover behaviour}
\label{sec:results_universal}

Since the discrete-time sensitivity envelope has the same generic functional form for both codes, it is natural to ask whether the crossover behaviour can be approximately described by a universal curve after suitable rescaling.

Normalizing at the numerically determined $t_{\mathrm{opt}}$, define
\begin{equation}
    x
    =
    \frac{T}{t_{\mathrm{opt}}},
    \qquad
    y_{\mathrm{opt}}
    =
    \sqrt{\tau t_{\mathrm{opt}}}.
    \label{eq:results_universal_variables}
\end{equation}
The corresponding rescaled crossover is
\begin{widetext}
\begin{equation}
    \Sigma_\tau^{U}(x)
    =
    \begin{cases}
        \displaystyle
        x
        \exp\!\left[
            D_{1/2}y_{\mathrm{opt}}
            (1-\sqrt{x})
            +
            D_1y_{\mathrm{opt}}^2
            (1-x)
        \right],
        & x\leq1,
        \\[10pt]
        \displaystyle
        \sqrt{\lfloor x\rfloor},
        & x>1.
    \end{cases}
    \label{eq:results_universal_curve}
\end{equation}
\end{widetext}
The coefficients $D_{1/2}$ and $D_1$ are code dependent and presented above in Sec.~\ref{sec:results_sensitivity} for the Shor and Steane codes.

\begin{figure*}[t]
    \centering

    \includegraphics[width=0.88\textwidth]{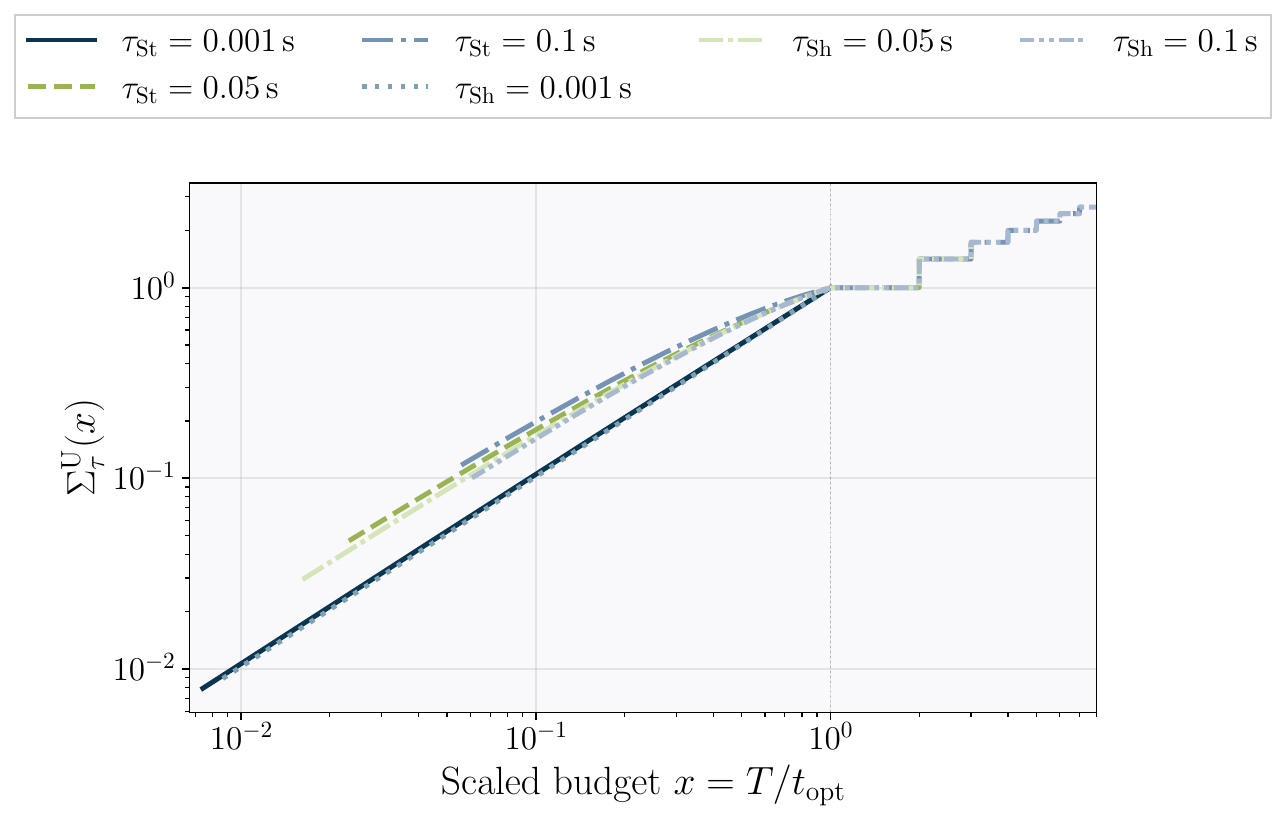}

    \caption{
        Rescaled discrete-QEC sensitivity. The sensing-time budget is
        normalized by the numerically determined
        $t_{\mathrm{opt}}$, and the sensitivity is normalized by its
        value at $t_{\mathrm{opt}}$.
    }
    \label{fig:universal_curve}
\end{figure*}

The rescaled curves are shown in Fig.~\ref{fig:universal_curve}. They do
not collapse exactly onto a single universal function, with deviations
becoming more pronounced as the correction interval is increased. This
behaviour reflects the simultaneous presence of the
$\sqrt{\tau t}$ and $\tau t$ decay contributions in the sensitivity
envelope. A further rescaling based on the dominant decay contribution
is developed in Appendix~\ref{app:universal_crossover}.

\subsection{Terminal-only recovery}
\label{sec:terminal_recovery}

To evaluate the effectiveness of a terminal-only variant of the protocol, a scheme is considered in which the intermediate recovery steps are omitted and a single recovery operation is applied immediately before the final measurement. The corresponding probability can be written as
\begin{equation}
    P_X^{\mathrm{term}}(\omega,t)
    =
    \operatorname{Tr}\!\left[
        \rho_0\,
        \mathcal E_t^\dagger
        \!\left(
            \mathcal R^\dagger(\Pi_X)
        \right)
    \right].
    \label{eq:terminal_probability}
\end{equation}

For the recovery map and the $X$-stabilizer projector considered here,
\begin{equation}
    \mathcal R^\dagger(\Pi_X)
    =
    \Pi_X.
    \label{eq:terminal_recovery_identity}
\end{equation}
It therefore follows that
\begin{equation}
    P_X^{\mathrm{term}}(\omega,t)
    =
    \operatorname{Tr}\!\left[
        \rho_0\,
        \mathcal E_t^\dagger(\Pi_X)
    \right]
    =
    P_X^{\mathrm{no\,QEC}}(\omega,t).
    \label{eq:terminal_equals_noqec}
\end{equation}

Thus, a recovery operation applied only at the final readout leaves both
the measurement probability and its frequency derivative unchanged in the case presented here. This shows that the terminal-only recovery variant does not improve the sensitivity relative to the corresponding uncorrected protocol.

\section{Discussion}

The results show that, within the protocol considered here, discrete-time
error correction can preserve Heisenberg-like temporal scaling over a
finite interrogation-time window when the signal and noise satisfy the
HNLS condition. In the present model, a local $Z$-type sensing
Hamiltonian is subject to independent $X$-type noise and protected using
partial quantum error correction based on a CSS code. As the correction
interval $\tau$ is reduced, the residual decay of the corrected signal is
suppressed, and the duration of the Heisenberg-like regime increases. For
any finite $\tau$, however, the remaining decay eventually leads to a
crossover to SQL-like scaling.

This finite-time regime is particularly relevant in realistic sensing
scenarios, where the available interrogation time is
limited.

The comparison between the Steane and Shor codes is considered both when
the HNLS condition is satisfied and when it is not. The
decay is code-dependent: the Steane projector contains
distinct non-oscillating and oscillating decay modes, whereas the
corresponding Shor contributions share the same decay rate.

The parallel-noise case illustrates the regime in which the HNLS
condition is not satisfied. For $Z$-type noise parallel to the signal,
the errors are not correctable within the present protocol. An encoded
probe without error correction may remain competitive over part of the
noise range, but it does not provide a metrological advantage over the
unencoded product-state benchmark. In contrast, for perpendicular
$X$-type noise, where the HNLS condition is satisfied and intermediate
error correction can be applied, the encoded protocols retain their
advantage over the GHZ benchmark for longer interrogation times. Their
performance relative to the independent-product probe improves as the
correction interval is reduced, while the advantage becomes less
pronounced for larger $\tau$.

The approximately code-independent form of the sensitivity envelope also
allows the HL-to-SQL crossover to be compared after rescaling by the
optimal interrogation time. The Steane and Shor results show similar
scaled behavior, although the collapse is not exact because both the
$\sqrt{\tau t}$ and $\tau t$ decay contributions remain relevant. When
one of these contributions dominates, the rescaled curves approach the
universal form derived in Appendix~\ref{app:universal_crossover}.

A terminal-only recovery does not improve the sensitivity for the chosen
$X$-stabilizer measurement. The advantage of the protocol therefore
is observed when recovery operations are applied during the sensing evolution
rather than from a correction performed only at readout. This conclusion
is specific to the present protocol, but the same reasoning applies
whenever the chosen readout is invariant under the adjoint recovery map.

The present construction relies explicitly on the CSS structure of the
code. The separation into $X$- and $Z$-type stabilizers allows
perpendicular Pauli noise to be corrected using one stabilizer sector,
while the complementary stabilizers are used to extract the sensing
signal. In the present protocol, $X$-type noise is corrected through
measurements of the $Z$-type stabilizers, while the $X$-stabilizer
projector provides the metrological readout for the $Z$-type signal. An
analogous construction can be obtained with the roles of $X$ and $Z$
interchanged.

The analytical formalism developed here applies to arbitrary CSS codes,
including higher-distance constructions. The code dependence enters
through the noisy evolution matrix $E^{(g)}$ and the recovery matrix $R$,
so the dependence on a single parameter such as the distance $d$ is not
isolated from the stabilizer and recovery structure. Nevertheless,
higher-distance codes are expected to extend the protected
interrogation-time window and increase $t_{\mathrm{opt}}$. Applying the
present framework to such codes would allow this dependence to be
quantified directly.

In the case of a small parallel noise component, it would remain uncorrected within this scheme
and would shorten the protected interrogation window. Extending the
method to more general noise models, or to stabilizer codes without the
same CSS separation, would require modified recovery and estimation
strategies.

Finally, the analysis assumes ideal syndrome measurements, negligible
recovery duration, and independent Markovian noise. Finite recovery
times, faulty syndrome extraction, correlated noise, or additional noise
introduced during correction would modify the effective decay rates and
could shift the optimal correction interval. Including these correction
costs would be expected to reduce the performance of the corrected
protocol relative to the idealized comparisons with the uncorrected GHZ
and product-state benchmarks. These effects provide a natural next step
toward assessing the practical performance of the protocol.

\section{Conclusion}

A quantum-metrology protocol based on CSS codes was studied for a
$Z$-type sensing Hamiltonian subject to parallel and perpendicular
noise. A general analytical framework was developed for describing the discrete-time corrected dynamics of an arbitrary CSS code within the present setting. The code dependence is captured by the noisy-evolution and recovery maps, from which the measurement probability, its frequency derivative, and the corresponding sensitivity can be obtained. The framework was evaluated explicitly for the Steane and Shor codes.

For parallel $Z$-type noise, the HNLS condition is not satisfied
and the relevant errors cannot be corrected within the present
protocol.

The main results concern perpendicular $X$-type noise. Discrete-time
error correction was shown to preserve Heisenberg-like temporal scaling
over a finite interrogation-time window, whose duration increases as the
correction interval $\tau$ is reduced. For any finite correction
interval, however, the remaining decay eventually leads to a finite
optimal interrogation time and a crossover to SQL-like scaling. A
recovery applied only at the final readout provides no advantage for the
chosen protocol, showing that intermediate correction steps are
essential.

The sensitivity envelope and its discrete-time decay were derived
analytically for general CSS codes and evaluated explicitly for the
Steane and Shor codes. The error-corrected encoded probes outperform the
GHZ benchmark over an extended interrogation-time range and, for
sufficiently small correction intervals, can also outperform the
independent-product-state benchmark.

Optimal interrogation times and the corresponding HL-to-SQL crossover
were determined. An approximately universal rescaling of the crossover
behavior was identified, although an exact collapse onto a single
code-independent curve was not obtained.

Future work should consider larger CSS codes, different sensing
Hamiltonians, more general noise models, and the effects of non-ideal
error correction, including finite recovery times and imperfect syndrome
measurements.

\section*{Code availability}

All code used to obtain the analytical and numerical results presented
in this work is available on
\href{https://github.com/Damuna/CSS-codes-for-Quantum-Metrology-with-Discrete-time-Error-Correction}{GitHub}.

\section*{Acknowledgements}

Most of this work was carried out at Leibniz University Hannover  with
support from the Collaborative Research Centre 1227 DQ-mat.

The authors gratefully acknowledge Prof.~Tobias Osborne for his guidance
and support during the development of this work.

Ugnė Liaubaitė completed her contribution to this work at Vilnius
University and acknowledges support from the Research Council of
Lithuania under the Program ``University Excellence Initiatives'' of the
Ministry of Education, Science and Sports of the Republic of Lithuania
(Measure No.~12-001-01-01-01, ``Improving the Research and Study
Environment''), Project No.~S-A-UEI-23-11. She also gratefully
acknowledges Prof.~Remigijus Paulavičius for his support at Vilnius
University.

\bibliography{references}

\clearpage
\onecolumngrid
\appendix

\section{Noisy evolution}
\label{appendix:noisy-stabilizer-evolution}

This appendix derives the analytical Heisenberg-picture evolution used
throughout the main text. The single-qubit dynamics are obtained first
for noise perpendicular and parallel to the sensing Hamiltonian. These
solutions are then extended to an arbitrary $X$-type stabilizer of a CSS
code and to the projector $\Pi_X$ used in the metrological readout.

\subsection{Single-qubit Heisenberg evolution}
\label{app:single_qubit_heisenberg}

We consider a single qubit evolving under
\begin{equation}
    H=\omega Z.
\end{equation}
For both noise models considered below, the Heisenberg evolution is
written as
\begin{align}
    X(t)
    &=
    a_\nu(t)X-b_\nu(t)Y,
    \label{eq:single_qubit_X_general}
    \\
    Y(t)
    &=
    b_\nu(t)X+c_\nu(t)Y,
    \qquad
    \nu\in\{X,Z\},
    \label{eq:single_qubit_Y_general}
\end{align}
where $\nu$ labels the noise model.

\subsubsection{$X$ noise}
\label{app:X_noise_single_qubit}

For noise perpendicular to the sensing Hamiltonian,
\begin{equation}
    L=\sqrt{\gamma}\,X.
\end{equation}
The adjoint Lindblad equation is
\begin{equation}
    \frac{dO}{dt}
    =
    \mathcal L^\dagger(O)
    =
    i[\omega Z,O]+\gamma(XOX-O).
    \label{eq:X_noise_lindblad}
\end{equation}
Its action on the Pauli basis is
\begin{align}
    \mathcal L^\dagger(I)&=0,\\
    \mathcal L^\dagger(X)&=-2\omega Y,\\
    \mathcal L^\dagger(Y)&=2\omega X-2\gamma Y,\\
    \mathcal L^\dagger(Z)&=-2\gamma Z.
\end{align}
Thus,
\begin{equation}
    \frac{d}{dt}
    \begin{pmatrix}
        X\\
        Y
    \end{pmatrix}
    =
    M_X
    \begin{pmatrix}
        X\\
        Y
    \end{pmatrix},
    \qquad
    M_X=
    \begin{pmatrix}
        0 & -2\omega\\
        2\omega & -2\gamma
    \end{pmatrix}.
    \label{eq:X_noise_matrix}
\end{equation}

The characteristic polynomial is
\begin{equation}
    \det(\lambda I-M_X)
    =
    \lambda^2+2\gamma\lambda+4\omega^2,
\end{equation}
and therefore, by the Cayley--Hamilton theorem,
\begin{equation}
    M_X^2+2\gamma M_X+4\omega^2 I=0.
    \label{eq:CH_X}
\end{equation}
Writing
\begin{equation}
    e^{M_Xt}=A_0(t)I+A_1(t)M_X
\end{equation}
gives
\begin{align}
    A_0'(t)&=-4\omega^2A_1(t),\\
    A_1'(t)&=A_0(t)-2\gamma A_1(t),
\end{align}
with $A_0(0)=1$ and $A_1(0)=0$.

Defining
\begin{equation}
    \Omega=\sqrt{4\omega^2-\gamma^2},
    \label{eq:Omega_definition}
\end{equation}
the solutions are
\begin{align}
    A_1(t)
    &=
    \frac{e^{-\gamma t}}{\Omega}
    \sin(\Omega t),
    \\
    A_0(t)
    &=
    e^{-\gamma t}
    \left[
        \cos(\Omega t)
        +
        \frac{\gamma}{\Omega}\sin(\Omega t)
    \right].
\end{align}
Hence,
\begin{equation}
    e^{M_Xt}
    =
    e^{-\gamma t}
    \begin{pmatrix}
        \displaystyle
        \cos(\Omega t)
        +\frac{\gamma}{\Omega}\sin(\Omega t)
        &
        \displaystyle
        -\frac{2\omega}{\Omega}\sin(\Omega t)
        \\[6pt]
        \displaystyle
        \frac{2\omega}{\Omega}\sin(\Omega t)
        &
        \displaystyle
        \cos(\Omega t)
        -\frac{\gamma}{\Omega}\sin(\Omega t)
    \end{pmatrix}.
    \label{eq:X_noise_exp}
\end{equation}

The coefficients in
Eqs.~\eqref{eq:single_qubit_X_general} and
\eqref{eq:single_qubit_Y_general} are therefore
\begin{align}
    a_X(t)
    &=
    e^{-\gamma t}
    \left[
        \cos(\Omega t)
        +
        \frac{\gamma}{\Omega}\sin(\Omega t)
    \right],
    \label{eq:aX}
    \\
    b_X(t)
    &=
    e^{-\gamma t}
    \frac{2\omega}{\Omega}\sin(\Omega t),
    \label{eq:bX}
    \\
    c_X(t)
    &=
    e^{-\gamma t}
    \left[
        \cos(\Omega t)
        -
        \frac{\gamma}{\Omega}\sin(\Omega t)
    \right].
    \label{eq:cX}
\end{align}

Using
\begin{equation}
    \partial_\omega\Omega=\frac{4\omega}{\Omega},
\end{equation}
their frequency derivatives are
\begin{align}
    \partial_\omega a_X(t)
    &=
    e^{-\gamma t}
    \frac{4\omega}{\Omega}
    \left[
        -t\sin(\Omega t)
        -
        \frac{\gamma}{\Omega^2}\sin(\Omega t)
        +
        \frac{\gamma t}{\Omega}\cos(\Omega t)
    \right],
    \label{eq:daX}
    \\
    \partial_\omega b_X(t)
    &=
    e^{-\gamma t}
    \left[
        \frac{8\omega^2t}{\Omega^2}\cos(\Omega t)
        -
        \frac{2\gamma^2}{\Omega^3}\sin(\Omega t)
    \right],
    \label{eq:dbX}
    \\
    \partial_\omega c_X(t)
    &=
    e^{-\gamma t}
    \frac{4\omega}{\Omega}
    \left[
        -t\sin(\Omega t)
        +
        \frac{\gamma}{\Omega^2}\sin(\Omega t)
        -
        \frac{\gamma t}{\Omega}\cos(\Omega t)
    \right].
    \label{eq:dcX}
\end{align}

The expressions above are written for $4\omega^2>\gamma^2$. For
$4\omega^2<\gamma^2$, they are analytically continued by defining
\begin{equation}
    \widetilde{\Omega}
    =
    \sqrt{\gamma^2-4\omega^2},
\end{equation}
such that $\Omega=i\widetilde{\Omega}$. The critical case
$4\omega^2=\gamma^2$ is obtained from the continuous limit
$\Omega\rightarrow0$.

\subsubsection{$Z$ noise}
\label{app:Z_noise_single_qubit}

For noise parallel to the sensing Hamiltonian,
\begin{equation}
    L=\sqrt{\gamma}\,Z.
\end{equation}
The adjoint Lindblad equation becomes
\begin{equation}
    \frac{dO}{dt}
    =
    \mathcal L^\dagger(O)
    =
    i[\omega Z,O]+\gamma(ZOZ-O).
    \label{eq:Z_noise_lindblad}
\end{equation}
The relevant Pauli operators satisfy
\begin{align}
    \mathcal L^\dagger(X)
    &=
    -2\gamma X-2\omega Y,
    \\
    \mathcal L^\dagger(Y)
    &=
    2\omega X-2\gamma Y.
\end{align}
Thus,
\begin{equation}
    \frac{d}{dt}
    \begin{pmatrix}
        X\\
        Y
    \end{pmatrix}
    =
    M_Z
    \begin{pmatrix}
        X\\
        Y
    \end{pmatrix},
    \qquad
    M_Z=
    \begin{pmatrix}
        -2\gamma & -2\omega\\
        2\omega & -2\gamma
    \end{pmatrix}.
    \label{eq:Z_noise_matrix}
\end{equation}

The characteristic polynomial gives
\begin{equation}
    M_Z^2
    +
    4\gamma M_Z
    +
    4(\gamma^2+\omega^2)I
    =
    0.
\end{equation}
Exponentiation yields
\begin{equation}
    e^{M_Zt}
    =
    e^{-2\gamma t}
    \begin{pmatrix}
        \cos(2\omega t) & -\sin(2\omega t)\\
        \sin(2\omega t) & \cos(2\omega t)
    \end{pmatrix}.
    \label{eq:Z_noise_exp}
\end{equation}
Hence,
\begin{align}
    a_Z(t)
    &=
    e^{-2\gamma t}\cos(2\omega t),
    \label{eq:aZ}
    \\
    b_Z(t)
    &=
    e^{-2\gamma t}\sin(2\omega t),
    \label{eq:bZ}
    \\
    c_Z(t)
    &=
    a_Z(t),
    \label{eq:cZ}
\end{align}
with
\begin{align}
    \partial_\omega a_Z(t)
    &=
    -2t\,b_Z(t),
    \label{eq:daZ}
    \\
    \partial_\omega b_Z(t)
    &=
    2t\,a_Z(t),
    \label{eq:dbZ}
    \\
    \partial_\omega c_Z(t)
    &=
    -2t\,b_Z(t).
    \label{eq:dcZ}
\end{align}

\subsubsection{Noiseless limit}

Setting $\gamma=0$ in either solution gives
\begin{align}
    a_0(t)&=\cos(2\omega t),\\
    b_0(t)&=\sin(2\omega t),\\
    c_0(t)&=\cos(2\omega t),
\end{align}
and therefore
\begin{align}
    X(t)
    &=
    \cos(2\omega t)X
    -
    \sin(2\omega t)Y,
    \\
    Y(t)
    &=
    \sin(2\omega t)X
    +
    \cos(2\omega t)Y.
\end{align}

\subsection{Noisy evolution of an $X$-type stabilizer}
\label{app:noisy_X_stabilizer}

Let $g\in G_X$ be an $X$-type stabilizer with binary support
$x_g\in\mathbb F_2^n$,
\begin{equation}
    g
    =
    X(x_g)
    =
    \prod_{j=1}^{n}
    X_j^{(x_g)_j},
    \qquad
    W_g=|x_g|.
    \label{eq:g_support}
\end{equation}
For either noise model,
\begin{equation}
    X(t)=a_\nu(t)X-b_\nu(t)Y,
    \qquad
    \nu\in\{X,Z\}.
\end{equation}
Since the Hamiltonian and noise act locally, the evolution factorizes
over the support of $g$,
\begin{equation}
    g(t)
    =
    \prod_{j=1}^{n}
    \left[
        a_\nu(t)X_j-b_\nu(t)Y_j
    \right]^{(x_g)_j}.
    \label{eq:g_factorized}
\end{equation}

Using $Y=iXZ$, the expansion can be labelled by a binary vector
$u\in\mathbb F_2^n$ whose support specifies the qubits on which the
$Y$ contribution is selected,
\begin{equation}
    g(t)
    =
    \sum_{\substack{
        u\in\mathbb F_2^n\\
        \operatorname{supp}(u)
        \subseteq
        \operatorname{supp}(x_g)
    }}
    a_\nu(t)^{W_g-|u|}
    \left[-ib_\nu(t)\right]^{|u|}
    gZ(u).
    \label{eq:g_general_expansion}
\end{equation}
with
\begin{equation}
    Z(u)
    =
    \prod_{j=1}^{n} Z_j^{u_j},
\end{equation}
and its Hamming weight by $|u|$.
For the logical-plus initial state,
\begin{equation}
    \bra{\psi_0}gZ(u)\ket{\psi_0}
    =
    \begin{cases}
        1, & Z(u)\in G_Z,\\
        0, & Z(u)\notin G_Z.
    \end{cases}
    \label{eq:logical_plus_Z}
\end{equation}
For the contributing terms, CSS commutation implies that $|u|$ is even.
Therefore,
\begin{equation}
    \langle g(t)\rangle_\nu
    =
    \sum_{\substack{
        u\in\mathbb F_2^n\\
        \operatorname{supp}(u)
        \subseteq
        \operatorname{supp}(x_g)\\
        Z(u)\in G_Z
    }}
    (-1)^{|u|/2}
    a_\nu(t)^{W_g-|u|}
    b_\nu(t)^{|u|}.
    \label{eq:g_expectation_general}
\end{equation}

For $Z$ noise,
\begin{align}
    \langle g(t)\rangle_Z
    &=
    e^{-2W_g\gamma t}
    \sum_{\substack{
        u\in\mathbb F_2^n\\
        \operatorname{supp}(u)
        \subseteq
        \operatorname{supp}(x_g)\\
        Z(u)\in G_Z
    }}
    (-1)^{|u|/2}
    \nonumber\\
    &\qquad\times
    \cos^{W_g-|u|}(2\omega t)
    \sin^{|u|}(2\omega t).
    \label{eq:g_Z_noise}
\end{align}

For uncorrected $X$ noise,
\begin{align}
    \langle g(t)\rangle_X
    &=
    e^{-W_g\gamma t}
    \sum_{\substack{
        u\in\mathbb F_2^n\\
        \operatorname{supp}(u)
        \subseteq
        \operatorname{supp}(x_g)\\
        Z(u)\in G_Z
    }}
    (-1)^{|u|/2}
    \nonumber\\
    &\qquad\times
    \left[
        \cos(\Omega t)
        +
        \frac{\gamma}{\Omega}\sin(\Omega t)
    \right]^{W_g-|u|}
    \left[
        \frac{2\omega}{\Omega}\sin(\Omega t)
    \right]^{|u|}.
    \label{eq:g_X_noise}
\end{align}

Finally,
\begin{align}
    \partial_\omega\langle g(t)\rangle_\nu
    &=
    \sum_{\substack{
        u\in\mathbb F_2^n\\
        \operatorname{supp}(u)
        \subseteq
        \operatorname{supp}(x_g)\\
        Z(u)\in G_Z
    }}
    (-1)^{|u|/2}
    \Big[
    \nonumber\\
    &\quad
    (W_g-|u|)
    a_\nu(t)^{W_g-|u|-1}
    b_\nu(t)^{|u|}
    \partial_\omega a_\nu(t)
    \nonumber\\
    &\quad+
    |u|
    a_\nu(t)^{W_g-|u|}
    b_\nu(t)^{|u|-1}
    \partial_\omega b_\nu(t)
    \Big].
    \label{eq:dg_general}
\end{align}

\subsection{Projector probability}
\label{app:projector_probability}

Since
\begin{equation}
    \Pi_X
    =
    \frac{1}{|G_X|}
    \sum_{g\in G_X}g,
\end{equation}
the projector probability follows directly from the individual
stabilizer expectation values,
\begin{equation}
    P_X(\omega,t)
    =
    \frac{1}{|G_X|}
    \left[
        1+
        \sum_{g\in G_X\setminus\{I\}}
        \langle g(t)\rangle
    \right].
    \label{eq:projector_probability_appendix}
\end{equation}
Similarly,
\begin{equation}
    \partial_\omega P_X(\omega,t)
    =
    \frac{1}{|G_X|}
    \sum_{g\in G_X\setminus\{I\}}
    \partial_\omega\langle g(t)\rangle.
    \label{eq:dprojector_probability_appendix}
\end{equation}

\section{Discrete-time corrected evolution}
\label{app:corrected_evolution}

We now consider perpendicular $X$ noise with recovery operations
separated by a fixed interval $\tau$. One correction cycle consists of
noisy evolution followed by recovery,
\begin{equation}
    \mathcal C_\tau
    =
    \mathcal R\circ\mathcal E_\tau,
    \qquad
    \mathcal C_\tau^\dagger
    =
    \mathcal E_\tau^\dagger
    \circ
    \mathcal R^\dagger.
    \label{eq:cycle_map}
\end{equation}

The noisy evolution of an $X$-type stabilizer generates operators of the
form
\begin{equation}
    O_u^{(g)}=gZ(u),
    \qquad
    u\in\mathbb F_2^n.
\end{equation}
Both the recovery and the noisy evolution preserve the span of these
operators, allowing the complete dynamics to be represented by finite
matrices.

\subsection{Recovery map}

Let $\{S_a^Z\}_{a=1}^{r_Z}$ be independent generators of the
$Z$-stabilizer group. A syndrome
$s=(s_1,\ldots,s_{r_Z})\in\mathbb F_2^{r_Z}$ is associated with
\begin{equation}
    \Pi_s
    =
    \prod_{a=1}^{r_Z}
    \frac{
        I+(-1)^{s_a}S_a^Z
    }{2},
    \label{eq:syndrome_projector}
\end{equation}
and with an $X$-type correction
\begin{equation}
    C_s
    =
    X(c_s)
    =
    \prod_{j=1}^{n}
    X_j^{(c_s)_j}.
    \label{eq:correction_operator}
\end{equation}

The recovery channel and its adjoint are
\begin{align}
    \mathcal R(\rho)
    &=
    \sum_s
    C_s\Pi_s\rho\Pi_s C_s^\dagger,
    \\
    \mathcal R^\dagger(O)
    &=
    \sum_s
    \Pi_sC_s^\dagger O C_s\Pi_s.
    \label{eq:recovery_channel}
\end{align}

For $O_u^{(g)}=gZ(u)$,
\begin{equation}
    \mathcal R^\dagger
    \left(
        O_u^{(g)}
    \right)
    =
    \sum_v
    R_{v,u}O_v^{(g)},
    \label{eq:recovery_matrix_action}
\end{equation}
where
\begin{equation}
    R_{v,u}
    =
    \frac{1}{2^n}
    \sum_s
    (-1)^{u\cdot c_s}
    \operatorname{Tr}
    \left[
        Z(u\oplus v)\Pi_s
    \right].
    \label{eq:recovery_matrix}
\end{equation}
Here $u\cdot c_s$ denotes the binary inner product modulo two. The
recovery matrix $R$ depends only on the code and the chosen recovery
rule and is independent of $\omega$.

\subsection{Noisy evolution between recovery operations}

Using $Y=iXZ$, the single-qubit evolution under $X$ noise can be
rewritten as
\begin{align}
    X
    &\longrightarrow
    a_X(\tau)X
    -
    i b_X(\tau)XZ,
    \\
    XZ
    &\longrightarrow
    -i b_X(\tau)X
    +
    c_X(\tau)XZ.
    \label{eq:local_X_noise_map}
\end{align}
Outside the support of $g$,
\begin{equation}
    Z
    \longrightarrow
    e^{-2\gamma\tau}Z.
\end{equation}
Consequently,
\begin{equation}
    \mathcal E_\tau^\dagger
    \left(
        O_u^{(g)}
    \right)
    =
    \sum_v
    E_{v,u}^{(g)}(\omega,\tau)
    O_v^{(g)}.
    \label{eq:noisy_matrix_action}
\end{equation}

For a qubit in the support of $g$, define
\begin{equation}
    f_j(v_j,u_j)
    =
    \begin{cases}
        a_X(\tau), & u_j=v_j=0,\\
        c_X(\tau), & u_j=v_j=1,\\
        -i b_X(\tau), & u_j\neq v_j.
    \end{cases}
    \label{eq:local_transition}
\end{equation}
The matrix elements can then be written compactly as
\begin{equation}
    E_{v,u}^{(g)}(\omega,\tau)
    =
    \prod_{j:(x_g)_j=1}
    f_j(v_j,u_j)
    \prod_{j:(x_g)_j=0}
    \delta_{u_j,v_j}
    e^{-2\gamma\tau u_j}.
    \label{eq:E_matrix}
\end{equation}
Their frequency derivatives follow directly by differentiating the
local factors using
Eqs.~\eqref{eq:daX}--\eqref{eq:dcX}.

\subsection{Evolution over multiple correction cycles}

Combining the recovery and noisy evolution, one complete correction
cycle maps the coefficient vector $\lambda$ as
\begin{equation}
    \lambda
    \longrightarrow
    A^{(g)}(\omega,\tau)\lambda,
    \qquad
    A^{(g)}(\omega,\tau)
    :=
    E^{(g)}(\omega,\tau)R.
    \label{eq:cycle_matrix}
\end{equation}

Let $e_0$ denote the basis vector corresponding to the all-zero binary
string. Since initially $g=gZ(0)$, after $M$ complete cycles
\begin{equation}
    \lambda^{(M)}
    =
    \left[
        A^{(g)}(\omega,\tau)
    \right]^M
    e_0.
    \label{eq:M_cycle_coeff}
\end{equation}

Define
\begin{equation}
    \chi_u
    =
    \begin{cases}
        1, & Z(u)\in G_Z,\\
        0, & Z(u)\notin G_Z.
    \end{cases}
\end{equation}
At a correction time $t=M\tau$,
\begin{equation}
    \langle g(t)\rangle_M
    =
    \chi^{\mathsf T}
    \left[
        A^{(g)}(\omega,\tau)
    \right]^M
    e_0.
    \label{eq:g_M_cycles}
\end{equation}

Since $R$ is independent of $\omega$,
\begin{align}
    \partial_\omega
    \left[
        A^{(g)}
    \right]^M
    &=
    \sum_{\ell=0}^{M-1}
    \left[
        A^{(g)}
    \right]^\ell
    \left(
        \partial_\omega E^{(g)}
    \right)
    R
    \left[
        A^{(g)}
    \right]^{M-1-\ell}.
    \label{eq:matrix_power_derivative}
\end{align}
Thus,
\begin{equation}
    \partial_\omega
    \langle g(t)\rangle_M
    =
    \chi^{\mathsf T}
    \partial_\omega
    \left[
        A^{(g)}
    \right]^M
    e_0.
    \label{eq:g_derivative_M}
\end{equation}

For a general running time $t$, define
\begin{equation}
    m(t)
    =
    \left\lfloor
        \frac{t}{\tau}
    \right\rfloor,
    \qquad
    r(t)
    =
    t-m(t)\tau,
    \qquad
    0\leq r(t)<\tau.
    \label{eq:general_time}
\end{equation}
There are then $m(t)$ complete correction cycles followed by a final
incomplete noisy interval. The coefficient vector is
\begin{equation}
    \lambda(t)
    =
    \left[
        A^{(g)}(\omega,\tau)
    \right]^{m(t)}
    E^{(g)}(\omega,r(t))
    e_0,
    \label{eq:general_time_coeff}
\end{equation}
and
\begin{equation}
    \langle g(t)\rangle
    =
    \chi^{\mathsf T}\lambda(t).
    \label{eq:g_general_time}
\end{equation}
The frequency derivative follows by applying the product rule to
Eq.~\eqref{eq:general_time_coeff}.

\section{Spectral reduction and sensitivity envelope}
\label{app:spectral_reduction}

\subsection{Krylov-space reduction}
\label{app:krylov_reduction}

For the projector probability, only the part of the operator space
reachable from $\Pi_X$ is required. Define
\begin{equation}
    \Pi_X^{(0)}
    =
    \Pi_X,
    \qquad
    \Pi_X^{(m)}
    =
    \left(
        \mathcal C_\tau^\dagger
    \right)^m
    \Pi_X.
\end{equation}
The relevant Krylov space is
\begin{equation}
    \mathcal K_{\Pi_X}
    =
    \operatorname{span}
    \left\{
        \Pi_X^{(0)},
        \Pi_X^{(1)},
        \Pi_X^{(2)},
        \ldots
    \right\}.
    \label{eq:krylov_space}
\end{equation}
Since the operator space is finite dimensional, this sequence eventually
closes. Let $K$ denote the representation of
$\mathcal C_\tau^\dagger$ on this reduced space.

If $K$ is diagonalizable, its eigenoperators satisfy
\begin{equation}
    \mathcal C_\tau^\dagger(V_j)
    =
    \lambda_j V_j.
\end{equation}
Expanding the measurement projector as
\begin{equation}
    \Pi_X
    =
    \sum_j c_j V_j,
\end{equation}
where $c_j$ is the expansion coefficient of $V_j$, gives, at
$t=M\tau$,
\begin{equation}
    P_X(t)
    =
    \sum_j A_j\lambda_j^M,
    \qquad
    A_j
    =
    c_j\operatorname{Tr}\!\left[\rho_0V_j\right].
    \label{eq:spectral_probability}
\end{equation}

For a real decaying mode $r$,
\begin{equation}
    A_r
    =
    c_r\operatorname{Tr}\!\left[\rho_0V_r\right].
\end{equation}

For a complex-conjugate pair,
\begin{equation}
    B_k
    =
    2\left|
        c_{k,+}
        \operatorname{Tr}\!\left[\rho_0V_{k,+}\right]
    \right|,
\end{equation}

\begin{equation}
    \phi_k
    =
    \arg\!\left[
        c_{k,+}
        \operatorname{Tr}\!\left[\rho_0V_{k,+}\right]
    \right].
\end{equation}
The stationary contribution is
\begin{equation}
    P_1
    =
    \sum_{\lambda_j=1}
    c_j\operatorname{Tr}\!\left[\rho_0V_j\right].
\end{equation}
Writing
\begin{equation}
    \lambda_j
    =
    |\lambda_j|
    e^{i\theta_j}
\end{equation}
defines the decay rate and oscillation frequency
\begin{equation}
    D_j
    =
    -\frac{1}{\tau}
    \log|\lambda_j|,
    \qquad
    \Omega_j
    =
    \frac{\arg\lambda_j}{\tau}.
    \label{eq:spectral_rates}
\end{equation}

Grouping real modes and complex-conjugate pairs gives
\begin{equation}
    P_X(t)
    =
    P_1
    +
    \sum_r
    A_r e^{-D_rt}
    +
    \sum_k
    B_k e^{-D_kt}
    \cos(\Omega_k t+\phi_k).
    \label{eq:spectral_form}
\end{equation}

If $K$ is not diagonalizable, the same modes are
accompanied by polynomial prefactors in $t$.

In the frequent-correction regime, the relevant eigenvalues approach
unity and can be expanded as
\begin{equation}
    \lambda_j(\tau)
    =
    1
    +
    \lambda_{j,1}\tau
    +
    \lambda_{j,2}\tau^2
    +
    \cdots.
    \label{eq:eigenvalue_expansion}
\end{equation}
Their coefficients are obtained order by order from
\begin{equation}
    \det
    \left[
        \lambda_j(\tau)I-K(\tau)
    \right]
    =
    0.
\end{equation}

\subsection{Sensitivity envelope}
\label{app:sensitivity_envelope}

For fixed $R_{\mathrm{rep}}$, the time dependence of the sensitivity is
determined by
\begin{equation}
    \delta\omega^{-1}(t)
    =
    \sqrt{R_{\mathrm{rep}}}
    \frac{
        |\partial_\omega P_X(\omega,t)|
    }{
        \sqrt{
            P_X(\omega,t)
            \left[
                1-P_X(\omega,t)
            \right]
        }
    }.
    \label{eq:app_sensitivity}
\end{equation}

In the frequent-correction regime, the dominant dependence on $\omega$
is carried by the oscillatory phases in
Eq.~\eqref{eq:spectral_form}, whereas the residual decay vanishes as
$\tau\rightarrow0$.

For the protocols considered here, the noiseless probability reaches
its maximum,
\begin{equation}
    P_X^{(0)}=1,
\end{equation}
at points
\begin{equation}
    \omega t=\kappa\pi,
    \qquad
    \kappa\in\mathbb Z.
\end{equation}
Close to such a point, define
\begin{equation}
    \delta
    =
    \omega t-\kappa\pi.
\end{equation}
The noiseless probability can then be expanded as
\begin{equation}
    P_X^{(0)}
    =
    1-a\delta^2+b\delta^4+\cdots.
\end{equation}

In the frequent-correction regime, the leading residual decay produces
a correction of order $\tau t$, such that
\begin{equation}
    P_X(t)
    \simeq
    1
    -
    a\delta^2
    +
    b\delta^4
    -
    c\tau t.
\end{equation}
Substitution into Eq.~\eqref{eq:app_sensitivity} gives the generic
behaviour
\begin{equation}
    \frac{
        \delta\omega^{-1}
    }{
        \delta\omega_{\mathrm{env},0}^{-1}
    }
    \simeq
    1
    -
    A\delta^2
    -
    B\frac{\tau t}{\delta^2}
    +
    \cdots,
    \label{eq:sensitivity_near_max}
\end{equation}
where $A,B>0$ and
$\delta\omega_{\mathrm{env},0}^{-1}$ denotes the noiseless sensitivity
envelope.

Optimizing with respect to $\delta$ gives
\begin{equation}
    \delta_{\mathrm{opt}}^4
    \simeq
    \frac{B}{A}\tau t,
\end{equation}
so that the leading correction to the optimized sensitivity scales as
$\sqrt{\tau t}$. The weak-decay sensitivity envelope can therefore be
written as
\begin{equation}
    \delta\omega_{\mathrm{env}}^{-1}(t)
    \simeq
    \delta\omega_{\mathrm{env},0}^{-1}(t)
    \exp
    \left[
        -D_{1/2}\sqrt{\tau t}
        -D_1\tau t
        +\cdots
    \right].
    \label{eq:sensitivity_envelope}
\end{equation}
The coefficients $D_{1/2}$ and $D_1$ are determined by the spectral
decay modes and are evaluated explicitly for the codes considered in
the main text.

\section{Code-specific analytical results}
\label{app:code_specific_results}

The general spectral construction derived in
Appendix~\ref{app:spectral_reduction} is now specialized to the
seven-qubit Steane code and the nine-qubit Shor code. For both codes,
the relevant eigenmodes of the reduced one-cycle map determine the
discrete-time decay of the projector probability and, consequently, the
sensitivity envelope.

\subsection{Steane code}
\label{app:steane_analytic}

For the Steane code, the $X$-stabilizer group contains eight elements,
\begin{equation}
    \Pi_X
    =
    \frac{1}{8}
    \sum_{g\in G_X} g.
    \label{eq:steane_projector}
\end{equation}
The seven non-identity $X$-stabilizers are equivalent under the
symmetries relevant to the present dynamics and have weight four.
It is therefore sufficient to consider one representative stabilizer
$g$.

\subsubsection{Probability evolution}

In the noiseless case, Eq.~\eqref{eq:g_expectation_general} gives
\begin{equation}
    \langle g(t)\rangle^{(0)}
    =
    \cos^4(2\omega t)
    +
    \sin^4(2\omega t)
    =
    \frac{3}{4}
    +
    \frac{1}{4}\cos(8\omega t).
    \label{eq:steane_g_noiseless}
\end{equation}
Consequently,
\begin{equation}
    P_{X,\mathrm{St}}^{(0)}(t)
    =
    \frac{1}{8}
    \left[
        1+7\langle g(t)\rangle^{(0)}
    \right]
    =
    \frac{25}{32}
    +
    \frac{7}{32}\cos(8\omega t).
    \label{eq:steane_P_noiseless}
\end{equation}

After diagonalizing the reduced map, only three non-trivial eigenvalues
contribute to $\langle g(t)\rangle$: one real eigenvalue $\lambda_r$
and one complex-conjugate pair $\lambda_\pm$. We write them as
\begin{equation}
    \lambda_r
    =
    e^{-D_r\tau},
    \qquad
    \lambda_\pm
    =
    e^{-D_o\tau}
    e^{\pm i\Omega_{\mathrm{St}}\tau}.
    \label{eq:steane_eigenvalues}
\end{equation}
In the frequent-correction regime, their decay rates are
\begin{align}
    D_r
    &=
    6\gamma^2\tau
    +
    \left(
        \frac{32}{3}\gamma\omega^2
        -
        32\gamma^3
    \right)\tau^2
    +
    O(\tau^3),
    \label{eq:steane_Dr}
    \\
    D_o
    &=
    18\gamma^2\tau
    +
    \left(
        \frac{32}{3}\gamma\omega^2
        -
        92\gamma^3
    \right)\tau^2
    +
    O(\tau^3),
    \label{eq:steane_Do}
\end{align}
while
\begin{equation}
    \Omega_{\mathrm{St}}
    =
    8\omega+O(\tau).
    \label{eq:steane_frequency}
\end{equation}

To leading order in the mode amplitudes, the stabilizer expectation value
therefore takes the form
\begin{equation}
    \langle g(t)\rangle
    \simeq
    \frac{3}{4}e^{-D_r t}
    +
    \frac{1}{4}e^{-D_o t}
    \cos\!\left(
        \Omega_{\mathrm{St}}t+\phi_{\mathrm{St}}
    \right),
    \label{eq:steane_g_spectral}
\end{equation}
and the corresponding projector probability is
\begin{equation}
    P_{X,\mathrm{St}}(t)
    \simeq
    \frac{1}{8}
    +
    \frac{21}{32}e^{-D_r t}
    +
    \frac{7}{32}e^{-D_o t}
    \cos\!\left(
        \Omega_{\mathrm{St}}t+\phi_{\mathrm{St}}
    \right).
    \label{eq:steane_P_spectral}
\end{equation}

\subsubsection{Sensitivity envelope}

Comparing Eq.~\eqref{eq:steane_P_spectral} with the single-frequency
form
\begin{equation}
    P_X=q+u\cos\theta,
\end{equation}
gives
\begin{equation}
    q(t)
    =
    \frac{1}{8}
    +
    \frac{21}{32}e^{-D_rt},
    \qquad
    u(t)
    =
    \frac{7}{32}e^{-D_ot}.
    \label{eq:steane_qu}
\end{equation}

Using the general sensitivity-envelope expansion derived in
Appendix~\ref{app:sensitivity_envelope}, the Steane decay coefficients
are
\begin{align}
    D_{1/2}^{\mathrm{St}}
    &=
    \frac{3}{4\sqrt{2}}
    \sqrt{
        \frac{3D_r+D_o}{\tau}
    },
    \label{eq:steane_Dhalf}
    \\
    D_1^{\mathrm{St}}
    &=
    \frac{D_o}{2\tau}.
    \label{eq:steane_Done}
\end{align}

Restoring the repetition factor used in the main text, the noiseless
sensitivity envelope is
\begin{equation}
    \delta\omega_{\mathrm{env},0}^{-1}
    =
    2\sqrt{7R_{\mathrm{rep}}}\,t,
    \label{eq:steane_env_noiseless}
\end{equation}
and the discrete-time envelope becomes
\begin{equation}
    \delta\omega_{\mathrm{St,env}}^{-1}(t)
    \simeq
    2\sqrt{7R_{\mathrm{rep}}}\,t
    \exp\!\left[
        -D_{1/2}^{\mathrm{St}}\sqrt{\tau t}
        -
        D_1^{\mathrm{St}}\tau t
    \right].
    \label{eq:steane_envelope}
\end{equation}

To leading order in $\tau$,
\begin{align}
    D_{1/2}^{\mathrm{St}}
    &=
    \frac{9}{2\sqrt{2}}\gamma
    +
    O(\tau),
    \\
    D_1^{\mathrm{St}}
    &=
    9\gamma^2
    +
    O(\tau).
    \label{eq:steane_envelope_leading}
\end{align}

\subsection{Shor code}
\label{app:shor_analytic}

For the Shor code, the $X$-stabilizer group contains four elements.
With two independent $X$-type generators,
\begin{equation}
    \Pi_X
    =
    \frac{1}{4}
    (I+S_1^X)(I+S_2^X)
    =
    \frac{1}{4}
    \left(
        I+S_1^X+S_2^X+S_1^XS_2^X
    \right).
    \label{eq:shor_projector}
\end{equation}
The three non-identity $X$-stabilizers are equivalent, and each acts on
two three-qubit blocks.

\subsubsection{Probability evolution}

For a single three-qubit block, the noiseless evolution gives
\begin{align}
    f_3^{(0)}(t)
    &=
    \cos^3(2\omega t)
    -
    3\cos(2\omega t)\sin^2(2\omega t)
    \nonumber\\
    &=
    \cos(6\omega t).
    \label{eq:shor_block_noiseless}
\end{align}
A nontrivial weight-six $X$-stabilizer contains two such blocks, and
therefore
\begin{equation}
    \langle S^X(t)\rangle^{(0)}
    =
    \cos^2(6\omega t)
    =
    \frac{1}{2}
    \left[
        1+\cos(12\omega t)
    \right].
\end{equation}
The noiseless projector probability is consequently
\begin{equation}
    P_{X,\mathrm{Sh}}^{(0)}(t)
    =
    \frac{5}{8}
    +
    \frac{3}{8}\cos(12\omega t).
    \label{eq:shor_P_noiseless}
\end{equation}

For one three-qubit block, the relevant reduced-map eigenvalues form a
complex-conjugate pair,
\begin{equation}
    \lambda_\pm
    =
    e^{-D_3\tau}
    e^{\pm i\Omega_3\tau},
\end{equation}
with
\begin{equation}
    D_3
    =
    3\gamma^2\tau
    +
    \left(
        8\gamma\omega^2
        -
        8\gamma^3
    \right)\tau^2
    +
    O(\tau^3),
    \label{eq:shor_D3}
\end{equation}
and
\begin{equation}
    \Omega_3
    =
    6\omega+O(\tau).
\end{equation}

For a weight-six stabilizer, the corresponding modes are
$\lambda_+\lambda_-$ and $\lambda_\pm^2$. All three have the same
modulus, giving a common decay rate
\begin{equation}
    D_{\mathrm{Sh}}
    =
    2D_3
    =
    6\gamma^2\tau
    +
    \left(
        16\gamma\omega^2
        -
        16\gamma^3
    \right)\tau^2
    +
    O(\tau^3),
    \label{eq:shor_D}
\end{equation}
while the oscillating pair has frequency
\begin{equation}
    \Omega_{\mathrm{Sh}}
    =
    12\omega+O(\tau).
    \label{eq:shor_frequency}
\end{equation}

The corresponding projector probability is therefore
\begin{equation}
    P_{X,\mathrm{Sh}}(t)
    \simeq
    \frac{1}{4}
    +
    \frac{3}{8}e^{-D_{\mathrm{Sh}}t}
    +
    \frac{3}{8}e^{-D_{\mathrm{Sh}}t}
    \cos\!\left(
        \Omega_{\mathrm{Sh}}t+\phi_{\mathrm{Sh}}
    \right).
    \label{eq:shor_P_spectral}
\end{equation}

\subsubsection{Sensitivity envelope}

Comparing Eq.~\eqref{eq:shor_P_spectral} with
$P_X=q+u\cos\theta$ gives
\begin{equation}
    q(t)
    =
    \frac{1}{4}
    +
    \frac{3}{8}e^{-D_{\mathrm{Sh}}t},
    \qquad
    u(t)
    =
    \frac{3}{8}e^{-D_{\mathrm{Sh}}t}.
    \label{eq:shor_qu}
\end{equation}
The corresponding envelope coefficients are
\begin{align}
    D_{1/2}^{\mathrm{Sh}}
    &=
    \frac{1}{2}
    \sqrt{
        \frac{D_{\mathrm{Sh}}}{\tau}
    },
    \label{eq:shor_Dhalf}
    \\
    D_1^{\mathrm{Sh}}
    &=
    \frac{D_{\mathrm{Sh}}}{2\tau}.
    \label{eq:shor_Done}
\end{align}

The noiseless sensitivity envelope is
\begin{equation}
    \delta\omega_{\mathrm{env},0}^{-1}
    =
    6\sqrt{3R_{\mathrm{rep}}}\,t,
    \label{eq:shor_env_noiseless}
\end{equation}
so that
\begin{equation}
    \delta\omega_{\mathrm{Sh,env}}^{-1}(t)
    \simeq
    6\sqrt{3R_{\mathrm{rep}}}\,t
    \exp\!\left[
        -D_{1/2}^{\mathrm{Sh}}\sqrt{\tau t}
        -
        D_1^{\mathrm{Sh}}\tau t
    \right].
    \label{eq:shor_envelope}
\end{equation}

To leading order in $\tau$,
\begin{align}
    D_{1/2}^{\mathrm{Sh}}
    &=
    \sqrt{\frac{3}{2}}\,\gamma
    +
    O(\tau),
    \\
    D_1^{\mathrm{Sh}}
    &=
    3\gamma^2
    +
    O(\tau).
    \label{eq:shor_envelope_leading}
\end{align}

\subsection{Universal crossover scaling}
\label{app:universal_crossover}

To compare the crossover behaviour for different correction intervals
and codes, a rescaled form of the sensitivity is introduced. If the
sensitivity envelope is dominated by a single decay contribution with
exponent $p\in\{1/2,1\}$, it can be written as
\begin{equation}
    \delta\omega_p^{-1}(t)
    =
    C t
    \exp\!\left[
        -D_p(\tau t)^p
    \right].
    \label{eq:single_mode_envelope}
\end{equation}

The corresponding continuous repeated-interrogation objective is
maximized at
\begin{equation}
    t_p
    =
    \frac{1}{\tau}
    \left(
        \frac{1}{2pD_p}
    \right)^{1/p},
    \label{eq:tp_general}
\end{equation}
or explicitly,
\begin{equation}
    t_p
    =
    \begin{cases}
        \displaystyle
        \frac{1}{D_{1/2}^{\,2}\tau},
        & p=\dfrac{1}{2},
        \\[10pt]
        \displaystyle
        \frac{1}{2D_1\tau},
        & p=1.
    \end{cases}
    \label{eq:tp_cases}
\end{equation}

Using $t_p$ as the crossover time, the sensitivity is written as
\begin{equation}
    \Sigma_\tau(t;t_p)
    :=
    \begin{cases}
        \delta\omega_{\mathrm{env}}^{-1}(t),
        & t\leq t_p,
        \\[8pt]
        \displaystyle
        \sqrt{\frac{t}{t_p}}\,
        \delta\omega_{\mathrm{env}}^{-1}(t_p),
        & t>t_p.
    \end{cases}
    \label{eq:crossover_function}
\end{equation}

The scaled time variable is defined as
\begin{equation}
    z
    =
    \left(
        \frac{t}{t_p}
    \right)^p,
    \label{eq:scaled_time}
\end{equation}
together with the repetition factor
\begin{equation}
    J(t;t_p)
    :=
    \begin{cases}
        1,
        & t\leq t_p,
        \\[6pt]
        \displaystyle
        \sqrt{\frac{t}{t_p}},
        & t>t_p.
    \end{cases}
    \label{eq:jump_factor}
\end{equation}

The rescaled crossover function is then defined as
\begin{equation}
    \Sigma_{\tau,p}^{U}(z)
    :=
    \left[
        \frac{
            \Sigma_\tau(t;t_p)
        }{
            \delta\omega_{\mathrm{env}}^{-1}(t_p)
            J(t;t_p)
        }
    \right]^p.
    \label{eq:universal_rescaling}
\end{equation}

For an exact single-term envelope,
Eq.~\eqref{eq:universal_rescaling} reduces to the universal form
\begin{equation}
    \Sigma_0^{U}(z)
    =
    \begin{cases}
        \displaystyle
        z\exp\!\left(
            \frac{1-z}{2}
        \right),
        & z\leq 1,
        \\[8pt]
        1,
        & z>1.
    \end{cases}
    \label{eq:universal_curve}
\end{equation}
The normalization by $J(t;t_p)$ removes the repetition factor and allows
the resulting curve to be independent of the dominant exponent $p$.

For the full two-term envelope, the dominant contribution is
identified from the logarithmic slope of the continuous
repeated-interrogation objective. Neglecting the integer constraint for
this comparison only, the relevant quantity is proportional to
\begin{equation}
    \frac{
        \delta\omega_{\mathrm{env}}^{-1}(t)
    }{
        \sqrt{t}
    }.
    \label{eq:continuous_repeated_objective}
\end{equation}
Its logarithmic slope is
\begin{equation}
    \frac{d}{d\log t}
    \log\!\left[
        \frac{
            \delta\omega_{\mathrm{env}}^{-1}(t)
        }{
            \sqrt{t}
        }
    \right]
    =
    \frac{1}{2}
    -
    \frac{1}{2}
    D_{1/2}\sqrt{\tau t}
    -
    D_1\tau t.
    \label{eq:envelope_log_slope}
\end{equation}

The contributions of the square-root and linear terms are therefore
\begin{equation}
    Q_{1/2}(t)
    =
    \frac{1}{2}
    D_{1/2}\sqrt{\tau t},
    \qquad
    Q_1(t)
    =
    D_1\tau t.
    \label{eq:slope_contributions}
\end{equation}

Their relative contribution is evaluated at the floor-aware numerical
optimum $t_{\mathrm{opt}}$ defined in
Eq.~\eqref{eq:topt},
\begin{equation}
    R_{\mathrm{slope}}
    :=
    \frac{
        Q_1(t_{\mathrm{opt}})
    }{
        Q_{1/2}(t_{\mathrm{opt}})
    }
    =
    \frac{
        2D_1\tau t_{\mathrm{opt}}
    }{
        D_{1/2}
        \sqrt{\tau t_{\mathrm{opt}}}
    }.
    \label{eq:Rslope}
\end{equation}
The dominant exponent is chosen according to
\begin{equation}
    p
    =
    \begin{cases}
        \dfrac{1}{2},
        & R_{\mathrm{slope}}<1,
        \\[6pt]
        1,
        & R_{\mathrm{slope}}\geq1.
    \end{cases}
    \label{eq:dominant_p}
\end{equation}

\begin{table}[t]
    \centering
    \caption{Ratio $R_{\mathrm{slope}}$, corresponding dominant exponent
    $p$, and associated crossover time $t_p$.}
    \label{tab:universal_scaling}
    \vspace{3mm}
    \setlength{\tabcolsep}{10pt}
    \begin{tabular}{lcccc}
        \toprule
        Code & $\tau$ [s] & $R_{\mathrm{slope}}$ & $p$ & $t_p$ [s] \\
        \midrule
        Steane & $10^{-3}$ & 0.01669 & $1/2$ & $8.158\times10^{3}$ \\
        Steane & 0.05      & 0.5121  & $1/2$ & 32.11 \\
        Steane & 0.10      & 1.25    & $1$   & 8.198 \\
        Shor   & $10^{-3}$ & 0.04056 & $1/2$ & $4.963\times10^{4}$ \\
        Shor   & 0.05      & 2.422   & $1$   & 18.76 \\
        Shor   & 0.10      & 7.247   & $1$   & 4.067 \\
        \bottomrule
    \end{tabular}
\end{table}

When one decay contribution dominates, the corresponding rescaled
crossover approaches the universal curve in
Eq.~\eqref{eq:universal_curve}. Deviations from this collapse quantify
the contribution of the subleading decay term.

\begin{figure}[t]
    \centering
    \includegraphics[width=0.88\columnwidth]
    {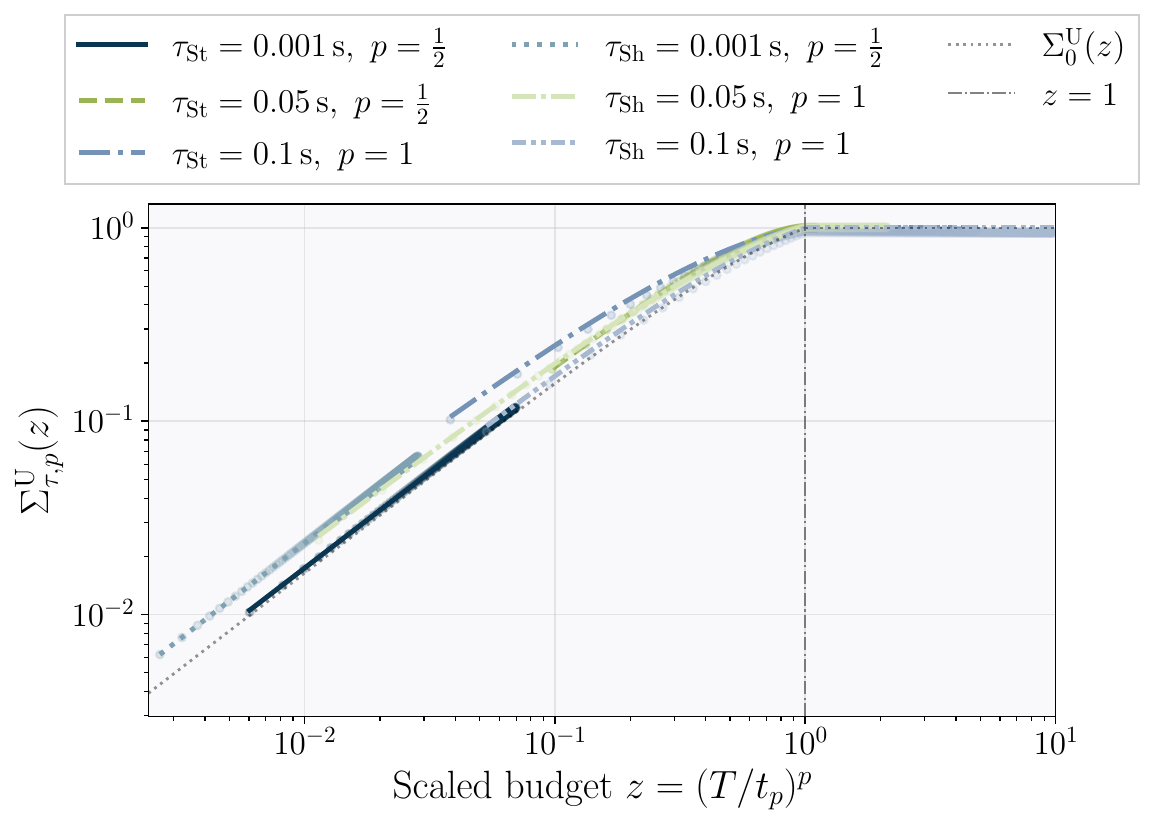}
    \caption{
        Universal scaling of the crossover function for the dominant
        exponent $p$. The rescaled quantity
        $\Sigma_{\tau,p}^{U}(z)$ is shown for the Steane and Shor codes
        at different correction intervals $\tau$. The reference curve
        $\Sigma_0^{U}(z)$ corresponds to the single-term prediction,
        and $z=1$ marks the crossover scale.
    }
    \label{fig:universal_crossover}
\end{figure}

\end{document}